\documentclass{article}

\pdfoutput=1

\usepackage{arxiv}

\usepackage[utf8]{inputenc} % allow utf-8 input
\usepackage[T1]{fontenc}    % use 8-bit T1 fonts
\usepackage{hyperref}       % hyperlinks
\usepackage{url}            % simple URL typesetting
\usepackage{booktabs}       % professional-quality tables
\usepackage{amsfonts}       % blackboard math symbols
\usepackage{nicefrac}       % compact symbols for 1/2, etc.
\usepackage{microtype}      % microtypography
\usepackage{lipsum}		% Can be removed after putting your text content
\usepackage{amsmath}
\usepackage{amssymb}
\usepackage{float}
\usepackage{graphicx}
\usepackage{natbib}
\usepackage{doi}

\title{Ranking earthquake forecasts using proper scoring rules: Binary events in a low probability environment}

\author{ \href{https://orcid.org/0000-0003-0154-6200
}{\includegraphics[scale=0.06]{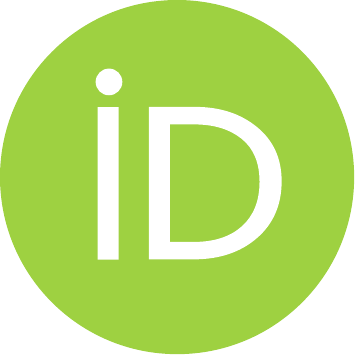}\hspace{1mm}Francesco Serafini} \\
	School of Geosciences\\
	University of Edinburgh\\
	Mayfield Rd, Edinburgh EH9 3JU\\
	\texttt{francesco.serafini@ed.ac.uk} \\
	\And
	Mark Naylor \\
	School of Geosciences\\
	University of Edinburgh\\
	Mayfield Rd, Edinburgh EH9 3JU\\
	\texttt{Mark.Naylor@ed.ac.uk} \\
	\And
	Maximilian Werner \\
	School of Earth Sciences Geophysics\\
	University of Bristol\\
	Bristol BS8 1TH\\
	\texttt{max.werner@bristol.ac.uk} \\
	\And
	Finn Lindgren \\
	School of Mathematics\\
	University of Edinburgh\\
	Mayfield Rd, Edinburgh EH9 3JU\\
	\texttt{Finn.Lindgren@ed.ac.uk} \\
	\And
	Ian Main \\
	School of Geosciences\\
	University of Edinburgh\\
	Mayfield Rd, Edinburgh EH9 3JU\\
	\texttt{Ian.Main@ed.ac.uk} \\
	
}

\renewcommand{\shorttitle}{Ranking earthquake forecasts using proper scoring rules}

\hypersetup{
pdftitle={Ranking earthquake forecasts using proper scoring rules: Binary events in a low probability environment},
pdfsubject={stat.AP},
pdfauthor={F.~Serafini, M.~Naylor, M.~Werner, F.~Lindgren, I.~Main},
pdfkeywords={Earthquakes forecasts comparison, Proper scoring rules, CSEP, Parimutuel gambling score},
}

\begin{document}
\maketitle

\begin{abstract}
	Operational earthquake forecasting for risk management and communication during seismic sequences depends on our ability to select an optimal forecasting model. To do this, we need to compare the performance of competing models in prospective experiments, and to rank their performance according to the outcome using a fair, reproducible, and reliable method, usually in a low-probability environment. The Collaboratory for the Study of Earthquake Predictability (CSEP) conducts prospective earthquake forecasting experiments around the globe. One metric employed to rank competing models is the Parimutuel Gambling score. We examine the suitability of this score for ranking earthquake forecasts. We prove analytically that this score is in general ‘improper’, meaning that, on average, it does not prefer the data generating model assuming this is known. In the special case where it is proper, we show it can still be used improperly. We compare its performance with two commonly-used proper scores (the Brier and logarithmic scores) using confidence intervals to account for the uncertainty around the observed score difference. We think that using confidence intervals enables a rigorous approach to distinguish between the predictive skills of candidate forecasts in addition to their rankings. Our analysis shows that the Parimutuel Gambling score is biased, and the direction of the bias depends on the forecasts taking part in the experiment. Our findings suggest the Parimutuel Gambling score should not be used to distinguishing between multiple competing forecasts, and for care to be taken in the case where only two are being compared.
\end{abstract}

% keywords can be removed

\section{Introduction}
\label{sec:intro}

Probabilistic earthquake forecasts are used to estimate the spatial and/or temporal evolution of seismicity and have potential utility during earthquake sequences, including those following notable earthquakes. For example, they have been applied to forecast (pseudoprospectively) the seismicity that followed the Darfield earthquake and in turn led to the 2011 Christchurch earthquake \citep{rhoades2016retrospective}, and to monitor induced seismicity at Groningen \citep{bourne2018exponential}. In Italy, earthquake probabilistic forecasts and ground-motion hazard forecasts are produced on a regular basis by the Instituto Nazionale di Geofisica e Vulcanologia (INGV) to inform the Italian government on the risk associated with natural hazard \citep{marzocchi2014establishment}. INGV is working to use probabilistic forecasts as a basis for modelling important quantities for operational loss forecasting such as the number of evacuated residents, the number of damaged infrastructure, the number of fatalities \citep{iervolino2015operational}. A wider uptake requires further demonstrations of the operational utility of the forecasts, and in presence of multiple alternative models, a fair and rigorous method to express a preference for a specific approach is needed. The Collaboratory for the Study of Earthquake Predictability (CSEP, see \cite{jordan2006brick,zechar2010csep,achievements}) is a global community initiative that seeks to make earthquake research more rigorous and open-science. This is done by comparing forecasts against future data in competition with those from other models through prospective testing in pre-defined testing regions. In this paper, we focus on comparing different forecasts that can be made from such competing models in the light of observed data.

In statistics, a common approach to compare probabilistic forecasts is the use of scoring rules \citep{gneiting2007strictly}. Scoring rules have been widely applied in many fields of science to measure the quality of a forecasting model and to rank competing models based on their consistency with the observed data and the degree of uncertainty around the forecast itself. Much of the underlying methodology and concepts (such as what it means to be a "good" forecast) have been developed for weather forecasts \citep{murphy1993good,jolliffe2003forecast}. A positively oriented scoring rule, to be effective, has to be \emph{proper}, which simply means that the highest score is achieved, on average, by the forecasting model "closer" to the distribution that has generated the observations. Various meaning of "closer" can be used depending on the context and the use that will be made of the forecasting model under evaluation, thus, a variety of proper scoring rules exists. Proper scoring rules are mathematically appealing for a range of different tasks: they can be used as utility function tailored to the problem at hand, they can be used as loss functions in parametric estimation problems and they can be used to rank competing models based on different aspects of the phenomenon under analysis \citep{rosen1996good,hyvarinen2005estimation,hernandez2012unified}.

CSEP aims to compare the predictive performance of diverse earthquake forecasts in a rigorous and reproducible fashion. The forecasts themselves are generated by underlying physical, stochastic or hybrid models using a variety of input data such as past seismicity, deformation rates, fault maps, etc \citep{field2014uniform, steacy2014new, bayliss2020data}. The two most widely used types are alarm-based forecasts and probabilistic forecasts. The first class of forecasts is usually expressed as a binary statement ("alarm" or "not alarm") based on the value of a precursory alarm function. In contrast, probabilistic forecasts, as intended in past CSEP experiments \citep{schor2007relm}, provide a distribution for the number of earthquakes. They can be expressed as grid-based forecasts (providing the expected number of events in each space-time-magnitude bin) or as catalogue-based (providing a number of simulated catalogues, \citeauthor{savran2020pseudoprospective},\citeyear{savran2020pseudoprospective} ). The forecasts are variously compared using a suite of community-endorsed tests. Depending upon the forecasts at hand, three common challenges are the need for a reference model, how to handle bins (or regions) for which the forecaster didn't provide a forecast and, the need to specify a likelihood. The latter has been partially solved by the possibility of considering a pseudo-likelihood \citep{savran2020pseudoprospective}.

Molchan diagrams \citep{zechar2008testing} and the area-skill score \citep{zechar2010area} do not need a likelihood and can be used to compare both alarm-based and probabilistic forecasts together. However, they need a reference model for assessing the significance of the results. This can be problematic because specifying a credible reference model is a difficult task \citep{stark1997earthquake,luen2008testing,marzocchi2011earthquake}. Likelihood-based tests  \citep{sch,zechar,rhoades2011efficient,schneider2014likelihood} allow for pairwise comparison without the need of a reference model, but can only be applied to probabilistic forecasts. Further, methods for grid-based forecasts rely on the Poisson assumption, which has been observed to be not realistic \citep{werner2008magnitude}. Moreover, pairwise comparison may lead to paradoxical results like model A is preferred to model B which is preferred to model C which is preferred to model A \citep{zechar2013regional}. Bayesian methods have been proposed \citep{marzocchi2012bayesian} but they also rely on the Poisson assumption. Catalogue-based forecasts, can be evaluated using a pseudo-likelihood approach \citep{savran2020pseudoprospective} which does not rely on the Poisson assumption and enable information gains and likelihood ratios to be used. However, the latter are unbounded and sensitive to low-probability events, meaning that they can be unduly influenced by a few observations \citep{holliday2005earthquake,zechar2014}. Lastly, in past experiments such as the Regional Earthquake Likelihood Models (RELM) \citep{field2007overview}, forecasters did not provide a forecast for all bins, some of them were left as missing value; for the methods outlined above, making a comparison is complex, given that considering only the overlapping space-time-magnitude volume may be too restrictive and introduce unfairness in the evaluation.

\cite{zhuang2010}, \cite{zechar2014} tried to overcome the difficulties outlined above by introducing the parimutuel gambling score, which provides a framework to evaluate different types of forecasts, with no need to explicitly specify a reference model or a likelihood, and with the ability to handle missing values in an intuitive way. This approach is based on the idea that alarm-based forecasts could be imagined as gamblers engaged in a game called the seismic roulette, where Nature controls the wheel \citep{main1997long,kossobokov2004earthquake,kossobokov2006testing}. In this framework, the forecasters are the gamblers, a forecast consists of a collection of probabilities for observable events (bets) like \emph{observing at least one earthquake in a specified space-time-magnitude bin}. Each bin represents a bet and the probability assigned by the forecaster represents the amount of money wagered. The observations consist of binary variables taking value 1 if the event occurs and zero otherwise. The forecaster gets a reward depending on the forecasted probability and the actual observation of an event or not. The forecasts are ranked based on their rewards. In this sense, the parimutuel gambling score is a positively oriented score (the higher, the better) for binary probabilistic forecasts. In this paper, we prove that the parimutuel gambling score is not proper in general but only in a specific situation and we compare its performance with two proper alternatives: the Brier \citep{brier1950verification} and the logarithmic \citep{goodj} score. 

To fairly compare the performance of the scores in a realistic framework, we use simulated data from a known model and we compare it with alternative models. In doing that, it is crucial to account for the uncertainty in the observed score difference. In fact, properness ensures that, at least on average, the scoring rule provides the correct ranking. However, the score calculated from any finite set of observations could be far from its average and, therefore, we need to account for uncertainty. In this paper, we show how to express a preference towards a model using confidence intervals for the expected score difference. This method introduces the possibility of not expressing a preference. Considering this outcome is potentially useful because it indicates that, for a scoring rule, the forecasts have similar performances, or the data are not enough to distinguish between models, or the bins' dimension is offset (too large or too small). 

In summary, the main goal of this article is to present the notion of proper scoring rules for probabilistic forecasts of binary events: why is it crucial for a scoring rule to be proper and what are the consequences of using an improper one? In Section 2 we define a proper and a strictly proper scoring rule, introduce the well known Brier and log scores as examples, and give a brief proof of their propriety. In Section 3, we introduce the parimutuel gambling score and analytically explore its improperness in the context of a forecast for a single bin. If a score is proper for single bins, then, the average score of different bins is also a proper score \citep{gneiting2007strictly}. In Section 4, we generalise to the case where we have multiple bins but with the same probability, \emph{Multiple Bins Single Probability}. This case is equivalent to considering the activity rate in each bin as independent and identically distributed. This is a significant assumption but allows us to calculate analytically the confidence intervals and the probability of expressing a preference for a given model. We generalise further to the case in which we have multiple bins but with a different probability for each bin, \emph{Multiple Bins Multiple Probabilities}. In this case, we do not have analytical results and we are required to use approximate confidence intervals and simulations to calculate the probability of expressing a preference. These simulations are now close to a real forecast scenario. We illustrate this case using simulations from  the time-independent 5 year adaptively-smoothed forecast for Italy \citep{werner2010adaptively}. We choose this model because the adaptively-smoothed approach performed well across multiple metrics in the RELM experiment \citep{zechar2013regional} and, as a result, was incorporated into the California seismic hazard map produced by the third Uniform California Earthquake Rupture Forecast model (UCERF3) \citep{field2014uniform}. 

\section{Proper Scores}
\label{sec:properscores}

Scoring rules quantify the quality of probabilistic forecasts, allowing them to be ranked. The quality depends on both the predictive distribution, produced by the model in true prospective mode, and on the subsequent observations. A scoring rule is a function of the forecast and the data measuring two factors: the consistency between predictions and observations and the sharpness of the prediction. Consistency assesses the calibration of the model, how well the forecast and the data agree, and is a joint property of the forecast and the data. Sharpness is a measure of the forecast uncertainty and is a property of the forecast only.  Different scoring rules measure the consistency and the sharpness of a forecast differently.  As in \cite{gneiting2007strictly}, we call $S(P\vert  x)$ the score for forecast $P$ given the observation $x$. In general, we use capital letters for random variables, lowercase letters for scalar quantities such as realizations of a random variable  (everything that is not random) and bold letters represents vectors. The only exception is $N$ which represents the number of bins. 

Thus, a scoring rule, given a forecast $P$, is a function of the observation only $S(P \vert  \cdot):\mathcal X \rightarrow [-\infty, \infty]$ where $\mathcal X$ is the set of all possible values of $x$. For consistency, we will use a \emph{positively orientated} convention, where a larger score indicates a better forecast. Assuming that the observations are samples from a random variable $X \in \mathcal X$ with true distribution $Q$, the score $S(P\vert  X)$ is a random variable itself, since it is a function of the random variable $X$. We define $S^E(P\vert Q)$ as the expected value of the scoring rule under the true distribution $Q$:

\begin{equation}
S^E(P\vert  Q) = \mathbb E_Q[S(P\vert X)] .
\label{eq:exp_value_score}    
\end{equation}

A positively oriented scoring rule $S$ is said to be \emph{proper} if, for any forecast $P$ and any true distribution $Q$, $S^E(Q\vert Q) \geq S^E(P\vert Q)$ holds. It is said to be \emph{strictly proper} if $S^E(Q\vert Q) = S^E(P\vert Q)$ if and only if $P = Q$. Propriety is essential, as it incentivises the assessor to be objective and to use the forecast $P$ "closer" to the true distribution $Q$. Different scoring rules rely on different meanings of closer. Also, proper scores can be used as loss functions in parameter estimation; in fact, since the likelihood assigned by a model to the observations can be seen as a proper scoring rule, the maximum likelihood estimator can be viewed as optimizing a score function \citep{huber1992robust}. Investigating the ability of a score of distinguishing between different instances of the same model (with different parameters values) may bring insight regarding parameters identifiability.

Here, we are interested in scoring rules for binary variables, in which the variable $X$ can be only 0 or 1, namely $X \in \{0,1\}$. Grid-based earthquake forecasts divide the region of interest into regular space-time-magnitude bins (e.g. the spatial region is divided in bins of $0.1 \times 0.1$ degrees, the magnitude by $0.1$ magnitude units, and the time is one $5$-year bin), and the forecasters estimate the expected number of earthquakes per bin. In this case, for example, the binary variable might be 0 for empty bins and 1 if at least one event occurs. The forecasts may be ranked based on the average score across different bins \citep{zechar2013regional}. 

Considering a single bin, for grid-based binary forecasts, where both the forecast $P$ and the true distribution $Q$ are specified by just one number: the probability of $X$ being 1. We call $p$ the probability assigned to the event $X = 1$ by the forecaster, and $p^*$ denotes the true probability. Thus, the expectation is given by
\begin{equation}
S^E(P\vert  Q) = S^E(p \vert  p^*) = p^*S(p\vert 1) + (1-p^*)S(p\vert 0) .
\label{eq:exp_score_deco}
\end{equation}

A scoring rule of this type is proper if, for any $p \in [0,1]$ and any $p \in [0,1]$, we have 
$$
S^E(Q\vert Q) \geq S^E(P\vert Q) .
$$

\noindent The properness of a score ensures that given two models $p_1, p_2$, the model with the greatest expected score $S^E(p_i\vert Q)$ is the closest to the true $p^*$. This notion can be generalized to rank a set of $k$ forecasts $p_1,...,p_k$ according to their expected scores.

Two of the most widely used strictly proper scoring rules, for binary data, are the Brier (or quadratic) score \citep{brier1950verification} and the logarithmic score \citep{goodj}. These are good candidates for evaluating this class of earthquake forecasts. Here we give the definitions of these two scores, including brief proofs of their propriety.

\subsection{ Brier Score }

The positively oriented Brier score \citep{brier1950verification} for a categorical variable $X$ (the binary case is obtained considering only two possible outcomes) can be defined by: 
\begin{equation}
S_B(P\vert x) = - \sum_{z\in \mathcal X} [p(z) - \mathbb{I}(z = x)]^2,
\label{eq:brier_def}
\end{equation}
where $\mathcal X$ is the set of possible outcomes, $p(z)$ is the forecasted probability of the event $X = z$, and $\mathbb I(z=x)$ is an indicator function assuming value 1 if $z=x$ and 0 otherwise. This definition differs from the original only in the sign, since the original Brier score is negatively oriented.  

The ordinary Brier score for binary events is the special case $\mathcal X=\{0,1\}$, with $p = p(1)$ and $1-p = p(0)$:
$$
\begin{aligned}
S_B(p\vert x) &= - [(1-p) - (1-x)]^2 - (p-x)^2 = -2(p-x)^2
&=
\begin{cases}
-2(p-1)^2, & x=1,\\
-2p^2, & x=0,
\end{cases}
\end{aligned}
$$
which has expectation
\begin{equation}
S^E_B(p\vert p^*) =
-2p^*(p-1)^2 - 2(1-p^*)p^2 
\label{eq:exp_brier}
\end{equation}
under the true event probability $p^*$. Taking the derivative with respect to $p$ and imposing it equal zero, we find that the value $p = p^*$ uniquely maximizes the function $S^E_B(p\vert p^*)$ which proves that the Brier score is strictly proper. 

\subsection{ Logarithmic Score }

The logarithmic (log) score for binary event forecasts is defined as
\begin{equation}
S_L(P\vert x) = \ln p_P(x) .
\label{eq:log_def}
\end{equation}

For $\mathcal X=\{0,1\}$, the expectation is
\begin{equation}
S^E_L(p\vert p^*) = p^* \ln(p) + (1-p^*) \ln(1-p)  ,
\label{eq:exp_log}
\end{equation}
which, once differentiated with respect to $p$ and set equal zero to identify the maximum, proves that also the log score is strictly proper.

\subsection{ Score Comparison }

Given an observation $x$, to express a preference between two forecasts $p_1$ and $p_2$, an important quantity is the score difference $\Delta$.
$$
\Delta(p_1,p_2,x) = S(p_1\vert x) - S(p_2\vert x) = 
\begin{cases}
S(p_1\vert 0) - S(p_2\vert 0) \quad \text{with prob} \quad 1 - p^*, \\
S(p_1\vert 1) - S(p_2\vert 1) \quad \text{with prob} \quad p^* .
\end{cases}
$$ 

For example, in the case of the Brier score we have
\begin{equation}
    \Delta_B(p_1,p_2,x)
    \begin{cases}
    -2(p_1^2 - p_2^2) &\quad \text{when} \quad x = 0,  \\
    -2[(1-p_1)^2 - (1-p_2)^2] &\quad \text{when} \quad x = 1,
    \end{cases}
    \label{eq:brier_diff}
\end{equation}
while in the case of the log score 
\begin{equation}
    \Delta_L(p_1,p_2,x)
    \begin{cases}
    \log(\frac{1-p_1}{1-p_2}) &\quad \text{when} \quad x = 0,  \\
    \log(\frac{p_1}{p_2}) &\quad \text{when} \quad x = 1.
    \end{cases}
    \label{eq:log_diff}
\end{equation}
In principle, if the expected value of $\Delta$ is positive we tend to prefer the first forecast, vice versa if it is negative. Considering the observation as a Bernoulli random variable $X \sim \text{Ber}(p^*)$, the difference $\Delta(p_1,p_2,X)$ is also a binary random variable, assuming the values $\Delta_0 = \Delta(p_1,p_2,0)$, $\Delta_1 = \Delta(p_1,p_2,1)$ with probabilities $1-p^*$ and $p^*$. The distribution of $\Delta(p_1, p_2, X)$ is therefore completely determined by the distribution of $X$:
\begin{equation}
\Delta(p_1, p_2, X) = X\Delta_1 + (1 - X)\Delta_0 = \Delta_0 + X(\Delta_1 - \Delta_0). 
\label{eq:delta_def}
\end{equation}
It follows that the expected value and variance of $\Delta(p_1, p_2, X)$ are determined by the properties of $X$:
\begin{align}
\mathbb E[\Delta(p_1, p_2, X)] &= \Delta_0 + \mathbb E[X](\Delta_1 - \Delta_0) = \Delta_0 + p^*(\Delta_1 - \Delta_0)
\label{eq:ex_value}
\\
\mathbb V[\Delta(p_1, p_2, X)] &= \mathbb V[X](\Delta_1 - \Delta_0)^2 = p^*(1- p^*)(\Delta_1 - \Delta_0)^2
\label{eq:variance}
\end{align}

We can give an alternative definition of the properness based on the random variable $\Delta(p_1,p_2,X)$. In fact, a scoring rule $S$ is said to be proper if $\mathbb E_X[\Delta(p, p^*, X)] \leq 0$ when $p \neq p^*$, no forecast have an expected score higher than the data generating model $p^*$. However, they can achieve the same score. $S$ is strictly proper if $\mathbb E_X[\Delta] = 0$ if and only if $p = p^*$, the highest score, on average, is achieved only by the data generating model. The definition implies, also, that proper scoring rules are invariant under linear transformations, in the sense that, a linear transformation of a proper score yields another proper score and the operation does not change the ranking.

\begin{figure}
% Use the relevant command to insert your figure file.
% For example, with the graphicx package use
  \includegraphics[width = 0.99\textwidth]{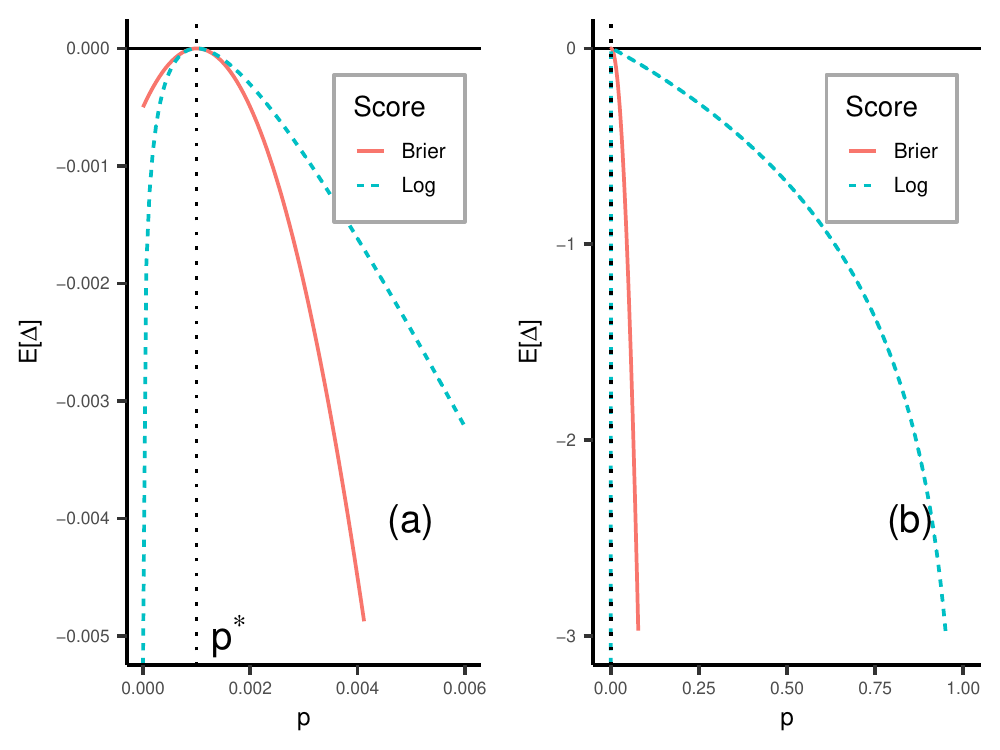}
% figure caption is below the figure
\caption{Differences in the scores expected value for a generic value of the forecast $p$ and the optimal forecast $p = p^*$, namely $\mathbb E[\Delta] = S^E(p\vert p^*) - S^E(p^*\vert p^*)$, in the case $p^* = 0.001$. Panel (a) $p \in (0,0.006]$, Panel (b) $p \in (0, 1)$. The expected Brier score have been normalised to match the curvature of the log score when $p = p^*$.}
\label{fig:1}       
\end{figure}

Figure \ref{fig:1} reports the expected score difference between a candidate forecast $p$ and the true value $p^* = 0.001$ using the Brier and the log score. The value $p^* = 0.001$ was chosen to be comparable to the estimated probability of having an event with magnitude greater than 5.5 calculated the days before the L'Aquila earthquake in the neighbourhood of where it struck [Fig. 4 in \cite{marzocchi2009real}]. To enable a visual comparison, the expected Brier score values have been normalized to match the curvature of the log score when $p = p^*$. This is done by multiplying the expected Brier score values by the ratio of the second derivatives of the two expected scores calculated at $p = p^*$. The proper score scale invariance ensures that the ranking obtained using the original and normalized version of the Brier score is unchanged. 

Both expected score differences are uniquely maximized at $p = p^*$ which means that the forecast matching the true probability has the highest expected score. This is an easy way to assess if a scoring rule for binary outcomes is proper or not. Furthermore, Figure \ref{fig:1} offers an example of how different scores penalize differently the same forecasts. The log score is asymmetric and takes into account the relative differences between the forecasts (equation \ref{eq:log_diff}), and if $p^* \neq 0$ the expected score for $p = \{0,1\}$ is $-\infty$. This means that the log score heavily penalizes overconfidence in the forecast. The Brier score, instead, considers the absolute difference between forecasts (equation \ref{eq:brier_diff}) resulting in a symmetric distribution. For example, using the Brier score, a forecast $p = 0$ will be preferred to any forecast in $(2p^*, 1)$, for any $p^* < 1/2$.

The choice of score, and consequently the style of penalty, should reflect the task at hand. Predicting $p = 0, 1$ means that we are absolutely certain about the outcome of $X$. If the forecasts under evaluation are planned to be used in an alarm based system, for which an alarm is broadcasted if the probability is above or below a certain threshold, being overconfident may put lives at risk and perhaps the log score would be the right choice in this situation. On the other hand, the Brier score may be suitable when being overconfident is less dangerous. This example illustrates the flexibility of proper scores and how important it is to choose the right one depending on the purposes of the forecast under evaluation.   

% Figure 1

\section{Improper scores}
\label{sec:improper}

Scores which are not proper are called improper. Being improper means that a model may exist with expected score greater than the data generating model. In the specific case of probabilistic forecasts for binary events, a score is improper if it is biased towards models which systematically under/overestimate the true probability $p^*$. In the context of earthquake forecasting experiments we do not know the true value of $p^*$. Therefore, it is crucial to use proper scoring rules for which we are sure that, at least on average, they will prefer the closest model to the data generating one. Improper scoring rules do not have this property, which implies that the smallest or the largest (or any other) forecast, on average, could achieve the highest score. This is in clear contrast with the aim of any forecasting experiment. Below, we demonstrate that the parimutuel gambling score \citep{zhuang2010, zechar2014} is an example of a scoring rule which is proper only in a specific situation and not in general.  

\subsection{Definition of the parimutuel gambling score}
\label{sec:defPG}

The parimutuel gambling score was designed to rank forecasting models for binary events and was applied to rank earthquake forecasting models in CSEP experiments \citep{taroni2018prospective, zechar2010risk}. Initially, it was used to compare models against a reference model  \citep{zhuang2010}, which is improper. Later, it was generalized to compare models against each other simultaneously \citep{zechar2014}, the case with only two players is the special case for which the score is proper, all the others are not. The score is based on a gambling scheme in which the forecasting models play the role of the gamblers and, for each observation, they obtain a reward proportional to the probability assigned by the gambler to the event occurring. In particular, it is a zero-sum game, in the sense that bids and rewards in each bin sum to zero, which makes the parimutuel gambling score relative to one forecast dependent on the other forecasts. 

In contract to the Brier and log scores, it is not possible to define the parimutuel gambling score using the form $S(p\vert x)$ because it needs at least two forecasts to be evaluated and is a function of them all. Given a set of $k$ forecasts $\mathbf p = (p_1,...,p_k)$, we define $S_G(\mathbf p\vert x)$ as the vector such that the $i$-th component, $S_{G,i}(\mathbf p\vert x)$, is given by the parimutuel gambling score of the $i$-th forecast, given $x$ has been observed. In the case of the Brier and log score the components of the vector $S(\mathbf p\vert x)$ are defined independently, in the case of the parimutuel gambling score they have to be defined jointly. Let $\bar p$  be the average probability involved in the gambling scheme, namely $\bar p = \sum_{i = 1}^k p_i/k$. The parimutuel gambling score relative to the $i$-th forecast is defined as
$$
S_{G,i}(\mathbf p\vert  x) = 
\begin{cases}
\frac{p_i}{\overline{p} } - 1, \quad & x = 1, \\
\frac{1 - p_i}{1 - \overline{p} } - 1, \quad & x = 0.
\end{cases}
$$
The above expression is a zero-sum game, meaning that $\sum_i S_{G,i}(\mathbf p\vert x) = 0$, therefore the rewards may be positive or negative. Each gambler obtains a positive reward if and only if they assign a greater probability to the observed event than the average gambler involved in the game. Vice versa, the reward is negative if the probability is smaller. 

The expected value with respect the true probability $p^*$ is given by
\begin{align}
S_{G,i}^E(\mathbf p\vert  p^*)  & = p^* \left( \frac{p_i}{\bar p} -1 \right) + (1 - p^*) \left( \frac{1-p_i}{1-\bar p} - 1 \right), \nonumber\\
& = \frac{p^*p_i}{\bar p} + \frac{(1 - p^*)(1 - p_i)}{1 - \bar p} - 1, \nonumber \\
& = \frac{(p_i - \bar p)\left( p^* - \bar p \right)}{\bar p(1 - \bar p)} .
\label{eq:exp_parimutuel}
\end{align}

The denominator involves all the probabilities in the game which demonstrates the interdependence with all other forecasts and complicates the study of the derivatives. However, it is still possible to prove that the gambling score is strictly proper when $k = 2$. In this case, $\mathbf p = (p_1,p_2)$, and
$$
\begin{aligned}
4 \bar p(1-\bar p) S_{G,1}^E(\mathbf p\vert  p^*) &=
4 \left(p_1 - \frac{p_1+p_2}{2}\right)\left( p^* - \frac{p_1+p_2}{2}\right),
\\
&= (p_1-p_2)\left( 2p^* - p_1-p_2\right), \\
&= -\left[(p_1-p^*)-(p_2-p^*)\right]\left[ (p_1-p^*)+(p_2-p^*)\right], \\
&= (p_2-p^*)^2 - (p_1-p^*)^2 .
\end{aligned}
$$
The expected reward of the first modeler is non-negative when $\vert p_1-p^*\vert \leq\vert p_2-p^*\vert $, implying that $p_1$ is favoured over $p_2$ if it is closer to the true probability $p^*$. In fact, if $p_2 = p^*$ then, $S_{G,1}^E(\mathbf p\vert p^*) \leq 0$, with the equality verified only for $p_1 = p^*$. Furthermore, the expected gambling score in this case is proportional to the expectation of the corresponding Brier score differences $\Delta_B=S_B^E(p_1\vert p^*)-S_B^E(p_2\vert p^*)$, thus, they produce, on average, the same rankings. 

\subsection{Improper use of proper score}

\begin{figure}
  \includegraphics[width = 0.99\textwidth]{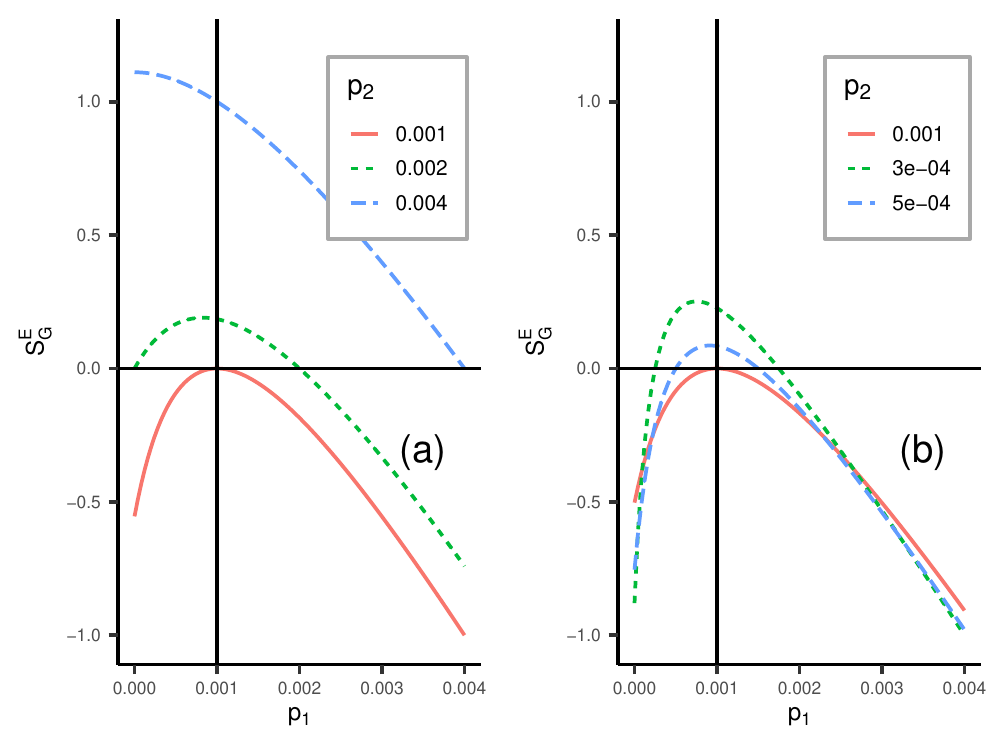}
\caption{ Expected value of the parimutuel gambling score (k = 2), $S^E_{G,1}(\mathbf p \vert  p^*)$, varying $p_1 \in (0,0.004) $, $p_2 = \{p^*, 2p^*, 4p^*$\} (a) and $p_2 = \{p^*, p^*/2, p^*/4\}$(b). The solid vertical line represents the true probability $p^* = 0.001$. The expected scores have been normalized so that their minimum is equal to -1. }
\label{fig:2}
\end{figure}

When comparing forecasting models, ensuring that the score is proper may not be sufficient. It also has to be used properly. The gambling score with $k = 2$ offers a nice example of this situation. We have demonstrated that the parimutuel gambling score is proper when $k = 2$, however, the dependence of the score value on all the forecasts involved in the comparison is a source of bias. In fact, $S_{G,1}^E(\mathbf p\vert  p^*) \geq S_{G,2}^E(\mathbf p\vert  p^*)$ when $p_1$ is closer to $p^*$ than $p_2$, however, $p_1 = p^*$ does not maximize $S_{G,i}^E(\mathbf p\vert  p^*)$ as shown in Figure \ref{fig:2}. This means that the score becomes biased when we rank forecasts based on the score difference against a reference model. 

% Figure 2

Formally, we are considering pairwise vectors $\mathbf p_1 = (p_1, p_0)$, $\mathbf p_2 = (p_2, p_0)$, etc., where $p_0$ is the reference model. For each of these we can estimate pairwise comparison score vectors $S_G(\mathbf p_1\vert x)$, $S_G(\mathbf p_2\vert x)$, and so on. The first component of each vector, namely $S_{G,1}(\mathbf p_1\vert x)$, $S_{G,1}(\mathbf p_2\vert x)$, etc, represents the score of $p_1$ and, respectively, $p_2$ against the reference model $p_0$. At this point, one would be tempted to rank the models based on $S_{G,1}(\mathbf p_1\vert x)$ and $S_{G,1}(\mathbf p_2\vert x)$, and this is the approach taken in \cite{taroni2014assessing} in which the official national time-independent model \citep{gruppo2004redazione} is used as reference model.

If the parimutuel gambling score is used to rank forecasts based on the score difference relative to a reference model, it will \emph{not} reliably favour the model closest to the true one, and the size of the bias will depend on the choice of the reference model. For example, in Figure \ref{fig:2}a, for $p_2 = 0.004$, the gambling score is maximized at $p_1 = 0$. This means that if the reference model is $p_0 = 0.004$, the overconfident forecast $p_1 = 0$ would be favoured by the ranking even if another forecast is perfect, e.g.\ $p_3 = p^*$. This problem can be particularly relevant in operational seismology where it is common for candidate forecasts to be compared against a reference model which is known to be based on simplistic assumptions (for example a homogeneous Poisson process). 

Hereon, the term pairwise gambling score refers to the comparison against a reference model as described in this section, while the term full gambling score will refer to the case where the forecasts compete directly against each other as we describe in the next section. Using this terminology, the full gambling score with $k = 2$ is the only proper score. 

\subsection{Improperness of the multi-forecast gambling score for $\boldsymbol{k \geq 3}$}

The generalized version of the full parimutuel gambling score, as presented in \cite{zechar2014}, for $k\geq 3$ is improper. For example, when $k = 3$ and $p_2 = p^*$, following equation \ref{eq:exp_parimutuel} the difference between the expected score for $p_1$ and $p_2$ is given by 
$$
\begin{aligned}
3\bar p(1-\bar p)[S^E_{G,1}(p_1, p^*, p_3\vert p^*) - S^E_{G,2}(p_1, p^*, p_3\vert p^*)] &= 3(p_1 - p^*)( p^* - \bar p)\\
&= (p_1 - p^*)(2p^* - p_1 - p_3 ) ,
\end{aligned}
$$
with both sides scaled by the common factor $3\bar p (1-\bar p)$. This means that when $2p^* - p_3 \geq p_1$, the forecast $p_1$ will have a positive score for any $p_1 \geq p^*$. Any value of $p_1 \in [p^*, 2p^* - p_3]$ will be preferred to $p_2$ that is equal to $p^*$. When $2p^* - p_3 \leq p_1$, with the same reasoning, $p_1$ is preferred over $p_2=p^*$ in the interval $[2p^* - p_3, p^*]$. 

\begin{figure}
  \includegraphics[width = 0.99\textwidth]{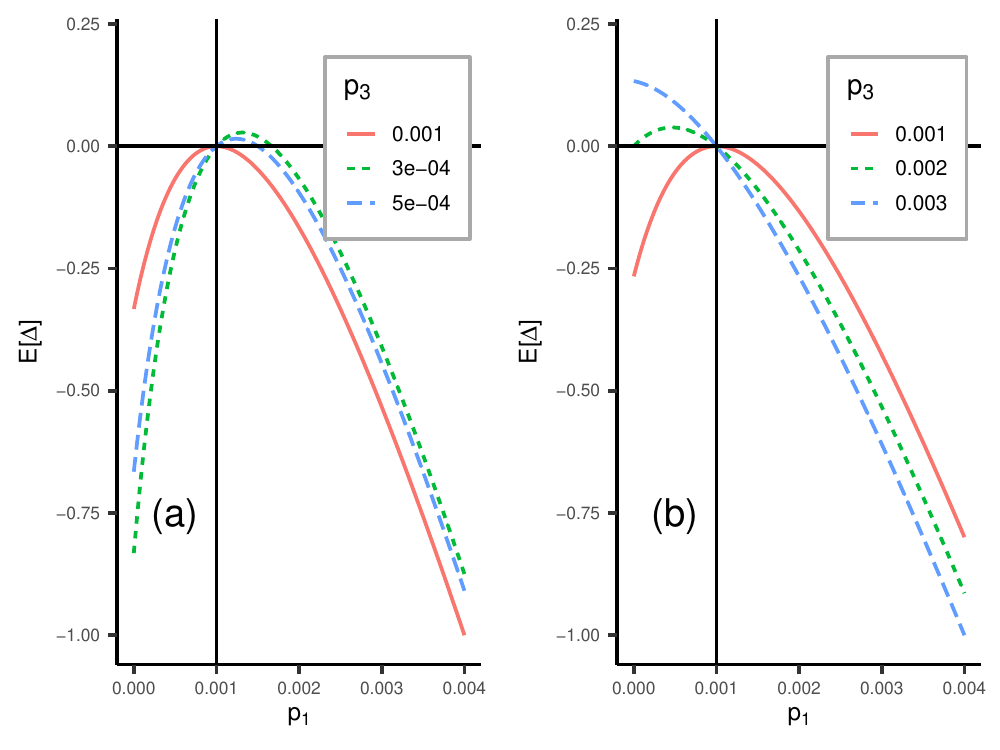}
\caption{Expected gambling score differences (k = 3) between $p_1$ and $p_2$, $S^E_{G,1}(\mathbf p\vert p^*) - S^E_{G,2}(\mathbf p\vert p^*)$, as a function of $p_1 \in (0,0.004)$, $p_2 = p^* = 0.001$ (vertical line), and  $p_3 \in \{p^*, p^*/2, p^*/3\}$ (a) and $p_3 \in \{p^*, 2p^*, 3p^*\}$ (b) .}
\label{fig:3}
\end{figure}

In Figure \ref{fig:3} we consider $k = 3$, $p^* = p_2 = 0.001$ and report the difference between the expected scores of $p_1$ and $p_2$, namely $S^E_{G,1}(\mathbf p\vert  p^*) - S^E_{G,2}(\mathbf p\vert p^*)$, for different values of $p_3$. The expected score difference is not maximize at $p_1 = p^*$, which means that the score is biased, and the "direction" of the bias depends on $p_3$ being greater than or equal to $p^*$.

% Figure 3

\begin{figure}
  \includegraphics[width = 0.99\textwidth]{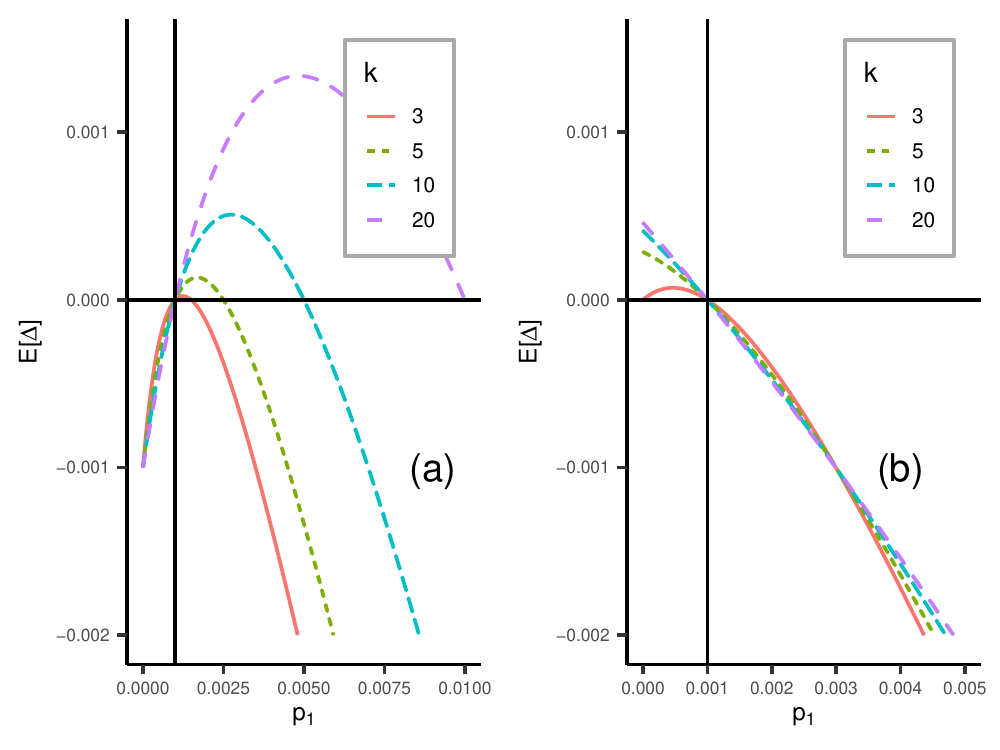}
\caption{Expected gambling score differences ($k \in \{3,5,10,20\}$) between $p_1$ and $p_2$ as a function of $p_1$ considering $p_2 = p^* = 0.001$ (vertical line). The averageof the forecast probabilities (excluding the first forecast) is the constant $\bar p_{-1} = p^*/2$ (a) and $\bar p_{-1} = 2p^*$ (b).}
\label{fig:4}
\end{figure}

Consider $k > 3$ gamblers who propose probabilities $\mathbf p = \{p_1,...,p_k\}$. It is helpful to consider the vector of probabilities that excludes the first component; we name this $\mathbf{p}_{-1} = \mathbf{p}/\{p_1\}$ and its mean $\bar p_{-1}$. Assuming $p_2 = p^*$, thus $\mathbf p = (p_1, p^*,...,p_k)$, we have that
$$
\begin{aligned}
k\bar p(1-\bar p)[S^E_{G,1}(\mathbf p\vert  p^*) - S^E_{G,2}(\mathbf p\vert p^*)]  &= k(p_1 - p^*)( p^* - \bar p),\\
&= (p_1 - p^*)[kp^* - p_1 - (k-1)\bar p_{-1}] ,
\end{aligned}
$$
from which we conclude that the expected score difference is positive when $p_1 \in [kp^* - (k-1)\bar p_{-1}, p^*]$ or $p_1 \in [p^*, kp^* - (k-1)\bar p_{-1}]$, depending on if $p^* \lessgtr \bar p_{-1}$. Specifically, when $p^* > \bar p_{-1}$ (Figure \ref{fig:3}a) the first gambler is encouraged to bet on a $p_1 > p^*$ and vice versa when $p^* < \bar p_{-1}$ (Figure \ref{fig:3}b). Furthermore, $p^*$ is always an extreme of the interval where the expected score difference is positive. Considering $\bar p_{-1}$ as fixed, the length of the interval is an increasing function of the number of gamblers $k$ (Figure \ref{fig:4}) which means that the size of the set of forecasts capable of obtaining a score value higher than the data generating model is an increasing function of the number of forecasts involved in the comparison.
% Figure 4
It is clear that the multi-forecast parimutuel gambling score favours models that are contrary to the average of the other forecasts. This could be particularly dangerous when evaluating the performance of earthquake forecasting models. For example, the trigger for an alarm being broadcast (or not) is often defined when the probability of having an earthquake above a certain magnitude exceeds a specified threshold. Using a model chosen looking at the full parimutuel gambling score could therefore lead to broadcasting alarms when they are not needed ($p\gg p^*$, 'crying wolf', Figure \ref{fig:4}a) or not broadcasting an alarm when needed ($p\ll p^*$, providing 'false reassurance', Figure \ref{fig:4}b).

The root of the problems with this score is that the score, relative to a candidate forecast, explicitly depends on the other forecasts. This design brings two problems: (i) the score, even in the special case when it is proper, can be used improperly and (ii) the score is never proper when considering more than two models. The Brier and log scores do not suffer from the same problem since the score of a forecast depends only on the forecast and the observation. Furthermore, the improperness demonstrated here can be expressed in terms that show that the gambling metaphor is part of the problem: If the outcome $x=0$ is likely (e.g.\ $p^* = 0.001$) and the majority of the forecasts have too large probabilities, then the expected gain is higher for an overconfident forecast, $p \ll p^*$, since that will give the forecaster a larger share of the total payout.

\section{ Forecasting across multiple bins }

Until now, we have analysed the expected score for a single bin, here we analyse the ability to express a preference between two forecasts using the average score across multiple bins. We assume to have access only to one observation $x_i \in \{0,1\}$ per bin . We analyse extensively the case in which the probability of observing $x_i = 1$,  is the same for each bin, $p^*_i = p^*$ for any $i = 1,...,N$. We refer to this as the Multiple Bins Single Probability case; the only quantity of interest is $p^*$ and a forecast is represented by a single value $p$. Even though, the Multiple Bins Single Probability case is clearly unrealistic in practice, it builds the basic concepts we will then use to explore the Multiple Bins Multiple Probabilities case where the probability of observing $x_i = 1$ is potentially different for each bin.

Considering multiple bins, we observe a realization of the random variable $X_i \sim \text{Ber}(p^*_i)$ for $i = 1,...,N$. A forecast is given by the vector $\mathbf p = (p_1,...,p_N)$ specifying the probability of $X_i = 1$ for each bin. Following the terminology in the literature regarding Bernoulli random variables, the event $X_i = 1$ is referred to as a \emph{success}. The quantity $X_S = \sum_i X_i$ is therefore referred as the sum of the observations or the number of successes or the number of active bins.

Given an arbitrary scoring rule $S(p\vert X)$, the average score associated with the forecast $\mathbf p$ is given by:
$$
S(\mathbf p\vert \mathbf X) = \frac{1}{N}\sum_{i = 1}^N S(p_i \vert  X_i).  
$$
The quantity $S(\mathbf p \vert  \mathbf X)$ is a random variable itself, because it is a function of random variables $\mathbf X$. To compare two forecasts $\mathbf p_1$ and $\mathbf p_2$, we study their score difference:
$$
\begin{aligned}
\Delta(\mathbf p_1, \mathbf p_2, \mathbf X) &= \frac{1}{N} \left(\sum_{i = 1}^N S(p_{1i} \vert  X_i) - \sum_{i = 1}^N S(p_{2i} \vert  X_i)\right), \\
&= \frac{1}{N}\sum_{i = 1}^N \Delta(p_{1i}, p_{2i}, X_i).
\end{aligned}
$$

The quantity $\Delta(\mathbf p_1, \mathbf p_2, \mathbf X)$ is also a random variable as it too depends on the vector of random variables $\mathbf X$. If $S(p\vert X)$ is a proper scoring rule, and if the expected value of the score difference is positive, namely $\mathbb E[\Delta(\mathbf p_1, \mathbf p_2, \mathbf X)] > 0$, the forecast $\mathbf p_1$ is "closer" to the true $\mathbf p^*$ than the alternative forecast $\mathbf p_2$. The expected value should be considered with respect the distribution of the observations $\mathbf X$. However, we do not observe the full distribution - we only observe a sample (i.e. we observe the quantity $\Delta(\mathbf p_1, \mathbf p_2, \mathbf x)$ which is a realization of the random variable $\Delta(\mathbf p_1, \mathbf p_2, \mathbf X)$). Even if the expected score difference $\mathbb E[\Delta(\mathbf p_1, \mathbf p_2, \mathbf X)]$ is positive, which means that we should express a preference for the first forecast, the observed score difference $\Delta(\mathbf p_1, \mathbf p_2, \mathbf x)$ may be negative and lead to the opposite conclusion. To avoid this problem we need to account for the uncertainty around the observed $\Delta(\mathbf p_1, \mathbf p_2, \mathbf x)$ which is the point estimate of the expected score difference $\mathbb E[\Delta(\mathbf p_1, \mathbf p_2, \mathbf X)]$. 

\subsection{ The distribution of score differences - Multiple Bins Single Probability }

For the Multiple Bins Single Probability case, the observation in each bin is a binary random variable $X_i \sim \text{Ber}(p^*)$, $i = 1,...,N$. Given an arbitrary scoring rule $S(p\vert X)$ and two candidate forecasts $p_1$ and $p_2$, the score difference for the $i$-th bin is a discrete random variable with distribution:
$$
\Delta(p_1, p_2, X_i) = 
\begin{cases}
\Delta_0 = S(p_1\vert 0) - S(p_2\vert 0) \quad\quad \text{with probability} \quad 1-p^* , \\
\Delta_1 = S(p_1\vert 1) - S(p_2\vert 1) \quad\quad \text{with probability} \quad p^*.
\end{cases}
$$

The forecasts are ranked based on the average score difference across all bins:
$$
\begin{aligned}
\Delta(\mathbf p_1, \mathbf p_2 , \mathbf X) &= \frac{1}{N}\sum_{i = 1}^N \Delta(p_1, p_2, X_i), \\
& = \frac{1}{N} \sum_{i = 1}^N\left( \Delta_0 + X_i(\Delta_1 - \Delta_0)\right),\\
& = \Delta_0 + \frac{X_S}{N}(\Delta_1 - \Delta_0),  
\end{aligned}
$$
where, $X_S = \sum_i X_i$ is the sum of all observations or, equivalently, the total number of successes. By definition, $X_S$ is the sum of $N$ (assumed to be) independent and identically distributed Bernoulli trials $X_i$. Therefore, $X_S$ has a Binomial distribution with size parameter $N$, the number of bins, and probability parameter $p^*$. When we observe a sample $x_1,...,x_N$, the observed score difference is given by:
$$
\Delta(\mathbf p_1, \mathbf p_2, \mathbf x) = \Delta_0 + \frac{\text x_S}{N}(\Delta_1 - \Delta_0),
$$
where $\text x_S$ is a realization of the random variable $X_S$. The observed score difference depends on the observations only through the quantity $\text x_S/N$. Thus, it is enough to study the quantity $\text x_S/N$ to make inference about the expected value of the score difference. The quantity $\text x_S/N$ it is said to be \emph{sufficient} \citep{fisher1922} with respect to the expected score difference because it contains all the information provided by the observations $x_1,...,x_N$ on the parameter of interest (in this case, the expected score difference $\mathbb E[\Delta(p_1, p_2, X)]$). For an introduction to statistical inference and the theory behind we refer to \citep{schervish2012,hastie2009elements}. 

\subsection{ Confidence Intervals for the Expected Score Difference }

A way to account for the uncertainty around the observed score difference is to consider an interval estimate of the expected value of the score difference. Once a sample $\mathbf x = x_1,...,x_N$ has been observed and the confidence interval calculated, if the entire interval lies above zero we express a preference towards $p_1$, alternatively if it lies below zero we express a preference towards $p_2$. If the interval contains the value zero we conclude that the observed sample does not contain enough information to express a preference. It is important to consider the latter case as a possible outcome because it is an indication that we need to collect more data or that the forecasts perform similarly (as measured by the score) and provide an additional information than the pure rankings. 

We are considering the confidence interval for the expected value of the score difference:
\begin{align}
\mathbb E[\Delta(p_1, p_2, \mathbf X)] &= \Delta_0 + \frac{\mathbb E[X_S]}{N}(\Delta_1 - \Delta_0), \nonumber\\ 
&= \Delta_0 + p^*(\Delta_1 - \Delta_0).
\label{eq:exp_delta_p}
\end{align}
Having an observation $\mathbf x = x_1,...,x_N$ per bin, the point estimate of $\mathbb E[\Delta(p_1, p_2, X)]$ is the observed score difference:
\begin{equation}
\Delta(p_1,p_2,\mathbf x) = \Delta_0 + \hat p (\Delta_1 - \Delta_0),
\label{eq:confint}
\end{equation}
where $\hat p = \text x_S/N$ is the observed probability of success. Comparing the equations \ref{eq:exp_delta_p} and \ref{eq:confint}, the point estimate of the score difference is retrieved plugging in the point estimate of the probability of success $\hat p$ in place of $p^*$. In the same way, to retrieve an interval estimate of the expected score difference is sufficient to retrieve an interval estimate of the probability of success $p^*$.  

Therefore, we need the confidence interval of level $\alpha$ for the true probability $p^*$ given observations $x_1,...,x_N$ from a Ber$(p^*)$, namely $CI_{p^*}(\alpha) = (\hat p_L, \hat p_U)$, and plug those values into expression \ref{eq:confint} to obtain a confidence interval for $\mathbb E[\Delta(p_1, p_2, X)]$, namely $CI_{\Delta}(\alpha) = (\Delta_L, \Delta_U)$. Various methods have been found to estimate $\hat p_L$ and $\hat p_U$, most of them relying on a Gaussian approximation. However, this approximation is not reliable for small sample sizes (number of bins $N$) and for values of $p^*$ close to zero or one, as in our case \citep{wallis2013binomial}. 

Hereafter, we use the Clopper-Pearson confidence interval \citep{clopper1934use}. This method is referred to as \emph{exact} because it relies on cumulative binomial probabilities rather than an approximation and is therefore more efficient and accurate than simulation based methods. The confidence interval with level $\alpha$ for $\hat p$ is given by:
\begin{align*}
p_L(\alpha) &= \text{BetaQ}(\frac{\alpha}{2}; \text x_S, N - \text x_S + 1 ), \\
p_U(\alpha) &= \text{BetaQ}(1 - \frac{\alpha}{2}; \text x_S + 1, N - \text x_S),
\end{align*}
where the function $\text{BetaQ}(q; a , b)$ is the $q$-th quantile of a Beta distribution with parameters $a$ and $b$. We can construct confidence intervals for $\mathbb E[\Delta(p_1, p_2, \mathbf X)]$ as follows:
\begin{align*}
\Delta_L &= \Delta_0 + \hat p_L(\Delta_1 - \Delta_0), \\
\Delta_U &= \Delta_0 + \hat p_U(\Delta_1 - \Delta_0) .
\end{align*}

The obtained confidence interval for $p^*$ depends on the data only through the sum of the observations $\text x_S$, which is a sufficient statistic for the problem. Similarly, the confidence interval for $\mathbb E[\Delta(p_1,p_2,\mathbf X)]$ depends on the data through the value of the sufficient statistic, $\text x_S$.  

\begin{figure}
  \includegraphics[width = 0.99\textwidth]{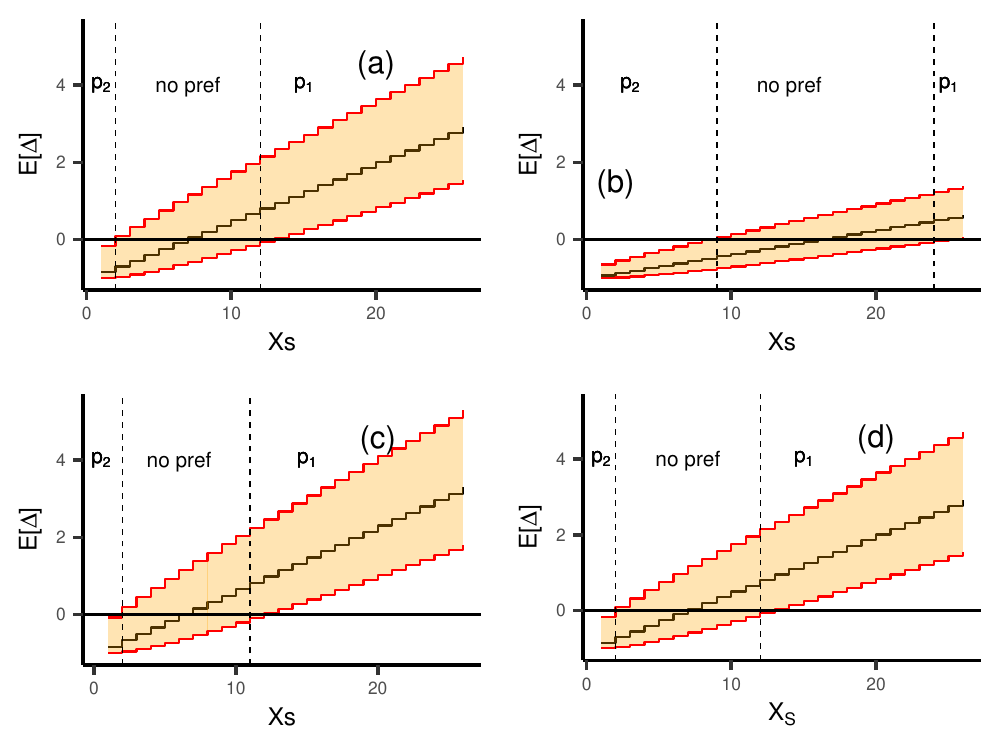}
\caption{Confidence interval (shaded area) and point estimate (black solid line) for $\mathbb E[\Delta]$ as a function of the number of observed successes $\text{x}_S$ considering $p_1 = 0.001$, $p_2 = p_1/3$, $p_0 = 5p_1$ and N = 10000. In each plot shows a different score: (a) Brier score; (b) pairwise gambling score; (c) logarithmic score; (d) full gambling score. Black solid line represents the observed score difference while the orange area represents the confidence interval. The black vertical dashed lines represent the interval of values of $\text{x}_S$ for which we do not express a preference}
\label{fig:5}
\end{figure}

Figure \ref{fig:5} shows the confidence interval for the score difference as a function of the sum of observations $x_S$ considering two competing forecasts $p_1 = 0.001$, $p_2 = p_1/3$, a reference model for the pairwise gambling score $p_0 = 5p_1$ and $N = 10,000$ bins. Here, we do not need to choose a value for $p^*$. Indeed, the confidence interval is determined solely by the forecast and observation. The Brier, log and full gambling score(Figure \ref{fig:5}a, \ref{fig:5}c, \ref{fig:5}d) all express a preference for $p_1$ if we observe $\text x_S > 12$, while they express a preference for $p_2$ when $\text x_S < 2$. This result is expected because $p_1 > p_2$, which means that $\text x_S > 12$ is much more probable under $p_1$ than $p_2$. In fact, the average number of successes using $p_1$ is $Np_1 = 10$ while $Np_2 = 3.34$. The same reasoning applies when we express a preference for $p_2$ ($\text x_S < 2$). 

The pairwise gambling score (Figure \ref{fig:5} (b)), instead, requires $\text x_S > 24$ to express a preference for $p_1$ and $\text x_S < 9$ to express a preference for $p_2$. It is heavily biased toward the forecast closer to zero. In fact, when $p_1$ is the true probability, the probability of observing $\text x_S > 24$ is less than 0.0001 and the probability of observing $\text x_S < 9$ is 0.33. Therefore, we are more likely to express a preference for $p_2$ than for $p_1$, even when $p_1 = p^*$. This reinforces the problems with employ improper scores introduced in Section \ref{sec:improper}. 

% Figure 5

\subsection{ Preference Probabilities } 

The confidence interval for the expected score difference, $CI_\Delta(\alpha)$, is a function of the competing forecasts, the scoring rule and depends on the data only through the sum of the observations $\text x_S = \sum_i x_i$. In particular, there are a range of values (between the dashed lines in Figure \ref{fig:5}) of $\text x_S$ for which we are not able to express a preference. We refer to this interval as $(\text x_{min}, \text x_{max})$. With respect to the sum of the observations $x_S$ there are only three possible outcomes:
\begin{align*}
\text x_S < \text x_{min} &\longrightarrow \text{preference for }p_2, \\
\text x_{max} \leq \text x_S \leq \text x_{max} &\longrightarrow \text{no preference}, \\
\text x_S > \text x_{max} &\longrightarrow \text{preference for }p_1.
\end{align*}

The values $\text x_{min}$ and $\text x_{max}$ are determined solely by $p_1$, $p_2$, the number of bins $N$ and the scoring rule. Table \ref{tab:1} reports the values of $\text x_{min}$ and $\text x_{max}$ for the scoring rules depicted in Figure \ref{fig:5}. These values can be used to compute the preference probabilities once a value for $p^*$ is assumed. Indeed, in the Multiple Bins Single Probability case, the distribution of $X_S$ is a Binomial distribution, $X_S \sim \text{Bin}(N, p^*)$. Table \ref{tab:2} reports the probabilities of i) no preference; ii) Preference for $p_1$; iii) Preference for $p_2$. The probabilities are calculated considering alternatively $p^*$ equal to $p_1$ (first half of the table) or $p_2$ (second half of the table). 

% Table 1
\begin{table}
\caption{Multiple Bins Single Probability case: table reporting the values $\text x_{min}$ and $\text x_{max}$ for the Brier, log, pairwise gambling (PG) and full gambling (FG) score. The reported values refers to the case where $N = 10,000$, $p_1 = 0.001$, $p_2 = p_1/3$ , and do not depend on $p^*$.}
\label{tab:1}       
% For LaTeX tables use
\begin{tabular}{l |  l  l}
\hline\noalign{\smallskip}
Score & $\text{x}_{min}$ & $\text{x}_{max}$  \\
\noalign{\smallskip}\hline\noalign{\smallskip}
Brier & 2 & 12 \\
Log & 2 & 11 \\
PG & 9 & 24 \\
FG & 2 & 12 \\
\noalign{\smallskip}\hline
\end{tabular}
\end{table}

The similarity among the values $\text x_{min}$ and $\text x_{max}$ for the proper scores lead to similar preference probabilities. The proper scores always assign the greatest probability to the case in which we are not able to express a preference, however, when $p^* = p_1$ it is unlikely to express a preference for $p_2$. Vice versa when $p^* = p_2$. There is a slightly difference between the Brier and the log score coming from the different penalty applied to forecasts close to zero. The log score penalises more heavily forecasts close to zero and, in fact, when $p_1 = p^*$ chances to express a preference for $p_1$ are higher than using the Brier score. The full gambling score for $p_1$ against $p_2$ is proportional to the Brier score difference between $p_1$ and $p_2$ (see Section 3.2) and thus, their preference probabilities coincide. Considering the pairwise gambling score the probability of expressing a preference for $p_1$ is always very close to zero, even when $p^* = p_1$. This shows again that it is possible to find a combination of $p_1$, $p_2$ and $p_0$ such that the model providing the smallest forecast obtains the highest reward with probability over $0.9$, even when the other forecast is equal to the true $p^*$. 

% Table 2
\begin{table}
\caption{Multiple Bins Single Probability case: table reporting for each score (row) the probabilities of expressing (or not) a preference using either the Brier, log-, pairwise gambling (PG) or full gambling (FG) score. The probabilities are calculated considering $N = 10,000$, $p_1 = 0.001$, $p_2 = p_1/3$ and considering two cases: $p^* = p_1$ and $p^* = p_2$.}
\label{tab:2}       
% For LaTeX tables use
\begin{tabular}{l |  l  l  l}
\hline\noalign{\smallskip}
Score & No pref & $p_1$ & $p_2$  \\
\noalign{\smallskip}\hline\noalign{\smallskip}
$p^* = p_1$ & & & \\
\noalign{\smallskip}\hline\noalign{\smallskip}
Brier & 0.7912 & 0.2083 & 0.0005 \\
Log & 0.6963 & 0.3032 & 0.0005 \\
PG & 0.6672 & 0.0000 & 0.3327 \\
FG & 0.7912 & 0.2083 & 0.0005 \\
\noalign{\smallskip}\hline\noalign{\smallskip}
$p^* = p_2$ & & & \\
\noalign{\smallskip}\hline\noalign{\smallskip}
Brier & 0.8454 & 0.0000 & 0.1545 \\
Log & 0.8453 & 0.0000 & 0.1545 \\
PG & 0.0073 & 0.2083 & 0.9927 \\
FG & 0.8454 & 0.0000 & 0.1545 \\
\noalign{\smallskip}\hline
\end{tabular}
\end{table}

\begin{figure}
  \includegraphics[width = 0.99\textwidth]{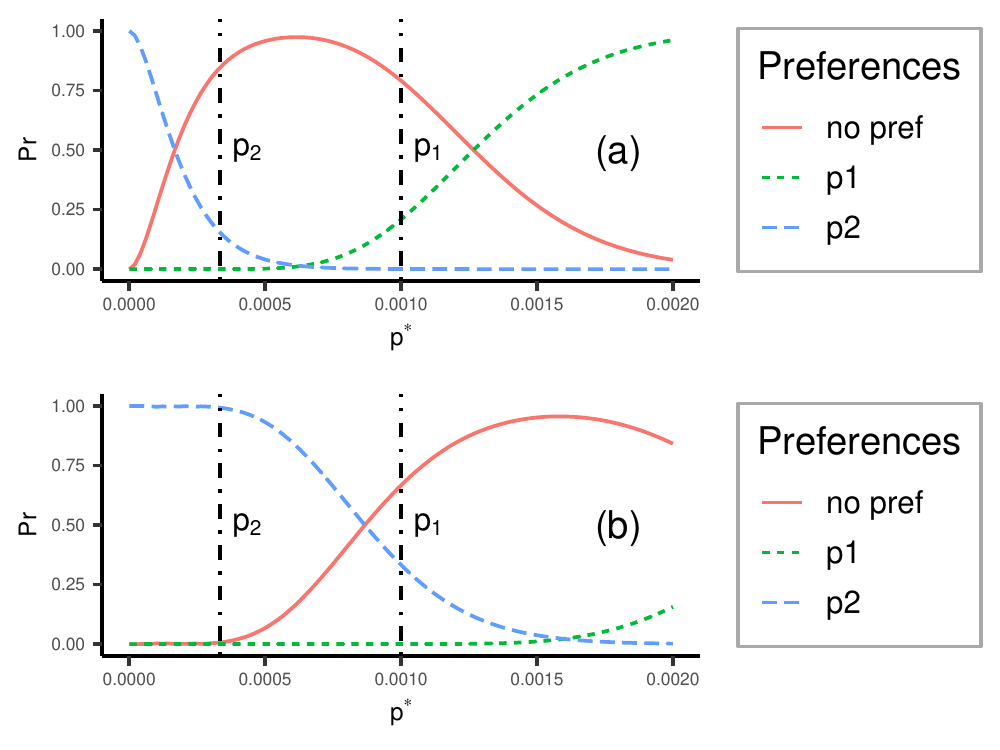}
\caption{Multiple Bins Single Probability case: each plot shows the probability of each possible outcome (solid lines no preference, dotted lines preference for $p_1$, dashed lines preference for $p_2$) as a function of $p^*$ using the Brier score (a) and the pairwise gambling score (b) considering $p_1 = 0.001$ and $p_2 = p_1/3$ (vertical lines), $p_0 = 5 p_1$ and $N = 10,000$. The true probability $p^*$ varies in $(10^{-6}, 2 \cdot 10^{-2})$ which is a realistic range of values in Italy.}
\label{fig:6}
\end{figure}

Figure \ref{fig:6} shows the preference probabilities as a function of $p^* \in (10^{-6}, 10^{-2})$ which is the range of values of the 5-year adaptively-smoothed forecast for Italy (aggregating over the magnitude bins) used later to illustrate the Multiple Bins Multiple Probabilities case. The Brier score behaves as expected. The probability of expressing a preference for $p_2$ increases as $p^*$ goes to zero, which is what we expect given $p_2 < p_1$. On the other hand, the probability of preferring $p_1$ increases when $p^*$ increases, because $p_1 > p_2$. Finally, the probability of not being able to express a preference is higher when $p_2 < p^* < p_1$ (Figure \ref{fig:6}a). The pairwise gambling score, instead, does not behave as expected. The probability of preferring $p_1$ is almost zero in the range of values of $p^*$ considered in the example. The two most probable outcomes are: expressing a preference for $p_2$ or not expressing a preference at all. 

% Figure 6

\subsection{ Probability of expressing a preference }

It is interesting to study how the probability for each case changes as a function of $p^*$ for different numbers of bins $N$ and different ratios between $p_1$ and $p_2$. To do that it is useful to focus only on two possible outcomes: expressing a preference and not expressing a preference. The probability of expressing a preference is given by the probability of observing a sample such that the sum of the observations $\text x_S$ is greater than $\text x_{max}$ or smaller than $\text x_{min}$. We refer to this probability as $\beta$ which is given by
$$
\beta = 1 - \Pr[\text x_{min} \leq X_S \leq \text x_{max}].
$$

This probability depends on the scoring rule, the forecasts $p_1$ and $p_2$,the number of bins $N$ and the true probability $p^*$. We study $\beta$ as a function of $p^*$ for different numbers of bins. In this artificial case, to increase the number of bins we are considering additional bins with the same probability, we are explicitly not splitting any bin; this is analogous to increase the data at hand applying the model to a larger spatio-temporal region.  

\begin{figure}
  \includegraphics[width = 0.99\textwidth]{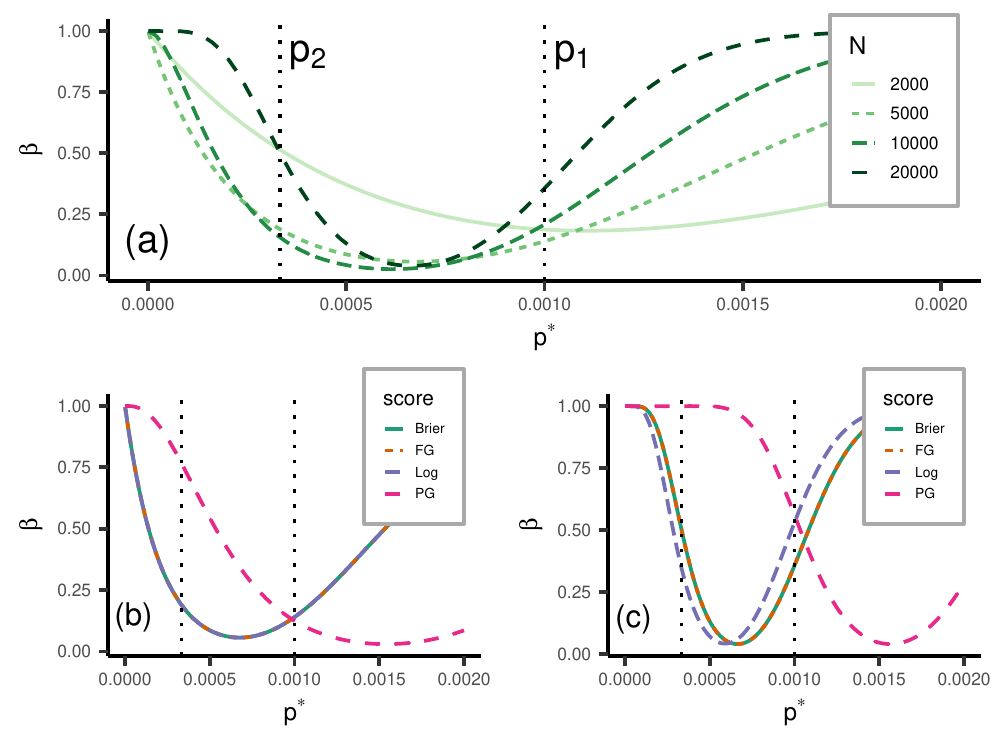}
\caption{Multiple Bins Single Probability case: (a) Brier score preference probability as a function of $p^*$ for different numbers of bins $N \in \{2000,5000,10000,20000\}$. (b-c) Probability of expressing a preference as a function of $p^*$. Colors represent the different scores: Brier, log , pairwise gambling (PG), and full gambling (FG) score. The Brier and FG scores coincide. The number of bins is fixed to N = 5000 (b) and N = 20000 (c).  We set $p_1 = 0.001$, $p_2 = p_1/3$ (vertical lines), $p_0 = 5p_1$, and $p^* \in (10^{-6}, 2^{-3})$ which is a realistic range of values in Italy.}
\label{fig:7}
\end{figure}

Figure \ref{fig:7}a considers only the Brier score. The region of $p^*$ presenting low values for $\beta$ shrinks when the number of bins increase which simply means that the more data we have, the more chances of expressing a preference. Moreover, $\beta$ is at the minimum when $p^* \in (p_2, p_1)$, which is reasonable because if the distances $\vert p^* - p_1\vert $ and $\vert p^* - p_2\vert $ are similar the probability of no preference should be high. The $N = 2000$ can be explained considering $p_1 = p^*$. In this case, the expected sum of observations is $Np_1 = 2$ and it is more probable to observe $X_S < 2$ than $X_S > 2$. Given that $Np_2 < Np_1$, the probability of not expressing a preference is high. 

Figure \ref{fig:7} presents the probability $\beta$ as function of $p^*$ for different scores with a fixed number of bins $N = 5000$ (b) and $N = 20000$ (c). For $N = 5000$, the proper scores (Brier, log and full gambling score) present the same values of $\beta$ for any value of $p^*$. For $N = 20000$, the proper scores start to behave differently. The Brier and full gambling score still coincide, while the log score is slightly different. Specifically, the log score presents higher $\beta$ values when $p^* = p_1$, and lower when $p^* = p_2$. This depends on the different penalties applied to forecasts close to zero. The log score presents greater chances of expressing a preference for $p_1$ when $p^* = p_1$ because the other forecast $p_2$ is smaller than $p_1$ and, therefore, penalized. On the other hand, when $p^* = p_2$ the log score presents smaller $\beta$ values than the Brier score.    

In contrast to the proper scores, the pairwise gambling score reaches its minimum $\beta$ value for $p^* > p_1$. Here, the pairwise gambling score tends to express a preference for the smaller forecast even when the other one is closer to $p^*$. This leads to higher values of $\beta$ when $p^* \in (p_2, p_1)$ because the pairwise gambling score will likely express a preference for $p_2$. Only when $p^* > p_1$ the probability of no preference grows and the value of $\beta$ decreases accordingly.

% Figure 7

Given that $p_1$ and $p_2$ are scalars, we can consider $\beta$ as a function of the ratio between $p_1$ and $p_2$,  $\omega = p_2/p_1$, for a fixed $p^*$. In principle, we expect that $\beta$ is an increasing function of $\omega$. We assume that the first forecast and the true probability are identical $p_1 = p^* = 0.001$. The reference model for the pairwise gambling score is $p_0 = 5p_1$ and we consider different numbers of bins $N \in \{2000, 5000, 10000, 20000\}$. The ratio $\omega = p_2/p_1$ varies in the interval $(0.1, 4)$. We expect low $\beta$ values when $\omega$ is around one (similar forecast) and high $\beta$ values otherwise.

\begin{figure}
  \includegraphics[width = 0.99\textwidth]{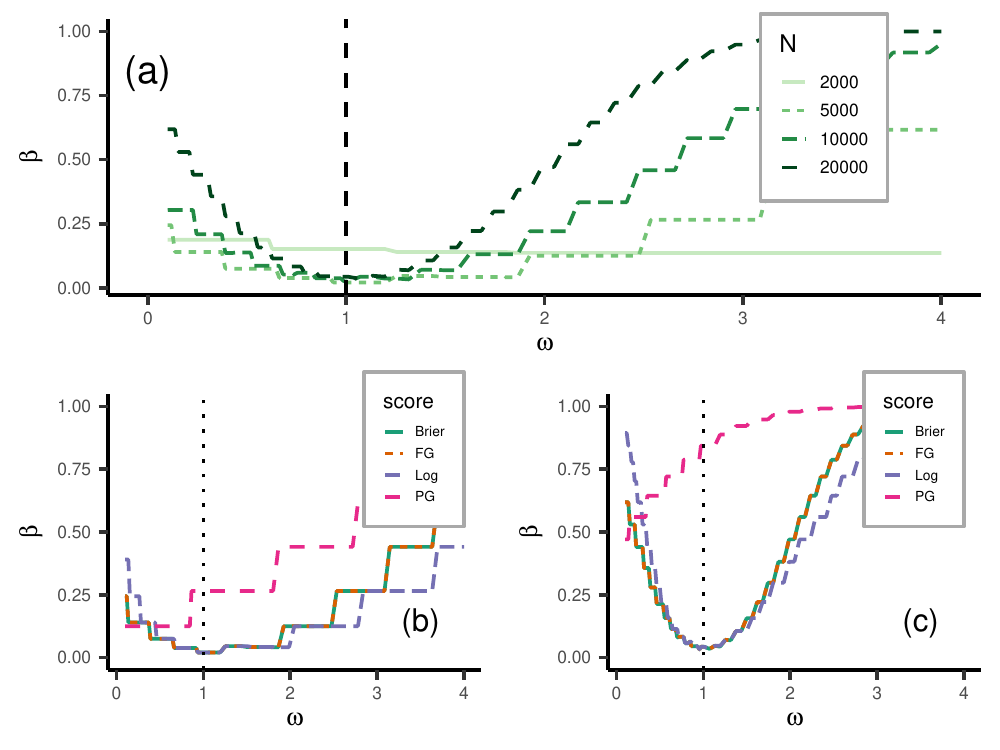}
\caption{Multiple Bins Single Probability case: (a) Probability of expressing a preference using the Brier score as a function of $\omega = p_2/p_1 \in (0.1,4)$  for different numbers of bins $N \in \{2000,5000,10000,20000\}$. We set $p_1 = p^* = 0.001$, and the reference model is $p_0 = 5p_1$. (Bottom) Probability of expressing a preference as a function of $\omega$. Colors represent the different scores: Brier, log, pairwise gambling (PG), and full gambling (FG) score. The number of bins is fixed to N = 5000 (b) and N = 20000 (c).}
\label{fig:8}
\end{figure}

Figure \ref{fig:8}a shows that, as expected, for $N > 2000$, $\beta$ has its minimum when $\omega = 1$. Considering $\omega$ as fixed, $\beta$ is an increasing function of the number of bins. Figure \ref{fig:8}b-c compares the $\beta$ values relative to different scores for a fixed number of bins, $N = 5000$ (b) and $N = 20000$ (c). The Brier and full gambling score coincide, whilst the log score presents slightly different $\beta$ values. As before, this is due to the different penalties applied to forecasts close to zero. 

The pairwise gambling score is not consistent with the trends in the proper scores. Using this score and considering $N = 20000$ (Figure \ref{fig:8}c), the probability $\beta$ is consistently greater than $0.5$ for the considered values of $\omega$. Considering that $p_1 = p^*$, the quantity $\omega$ is also the ratio between $p_2$ and $p^*$. This implies that regardless of $\omega$, we will erroneously express a preference for $p_2$ with a probability above $0.5$.  

Importantly, these sanity checks of a proposed scoring procedure can be done before looking at the observations. It is possible to check if forecasts can, in principle, be distinguished in light of the amount of expected data. We recommend the use of such exploitative figures when introducing a new scoring rule whose performance have not been tested. If the proposed scoring rule does not behave acceptably in this simple scenario, it is unlikely that it would behave acceptably in a real application.

% Figure 8

\subsection{ Score difference distribution - Multiple Bins Multiple Probabilities }

The Multiple Bins Multiple Probabilities case generalizes the Multiple Bins Single Probability case, and is much more similar to a real earthquake forecasting experiment. For example, the forecasts involved in the first CSEP experiments \citep{field2007overview, schorlemmer2007relm, zechar2013regional, michael2018preface} were mostly grid-based forecasts providing for each space-time-magnitude bin, the expected number of earthquakes. Then, the number of events in each bin is modelled using a Poisson distribution with intensity equal to the number of events provided by the forecasts and the probability of observing at least one event is calculated accordingly. In this scenario, we do not have analytical results for the score difference distribution and we need to recur to simulations.

\begin{figure}
  \includegraphics[width = 0.99\textwidth]{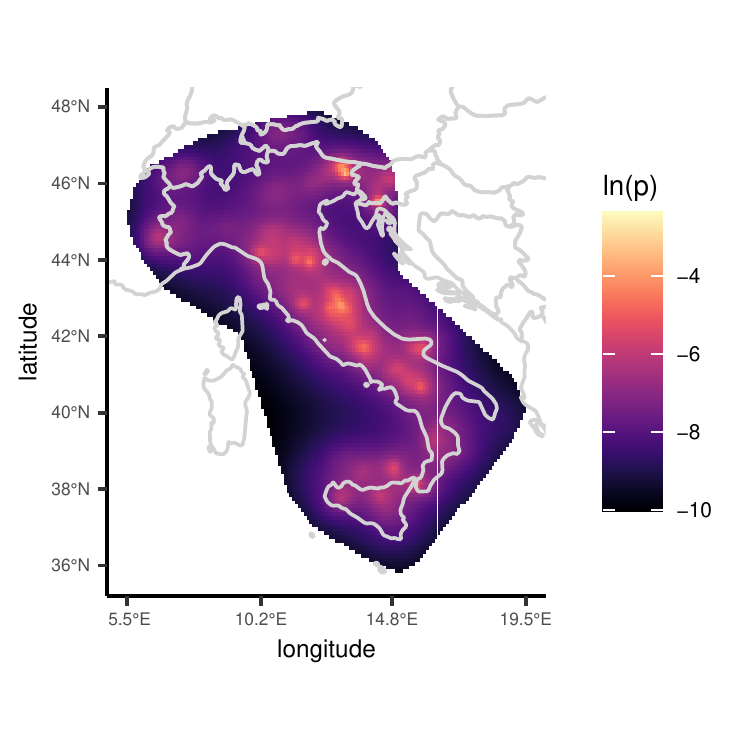}
\caption{5-year adaptively-smoothed forecast for Italy \citep{werner2010adaptively}. The figure shows for each spatial bin the natural logarithm of the probability of observing at least one earthquake at or above magnitude 4.95 in the period from January 1, 2010, to December 31, 2014.}
\label{fig:9}
\end{figure}

We now want to specify a true model which has more realistic probabilities. Since we do not actually know these in reality, we choose to work with one of the CSEP models that was submitted to the 2010 Italy experiment \citep{taroni2018prospective}. We choose to simulate synthetic data from the 5-year adaptively-smoothed forecast for Italy \citep{werner2010adaptively} and explore the ability of the scoring rules to discriminate between linearly scaled versions of this true model. This means that we are considering only one time bin of size 5 years, while the space-magnitude domain is divided in multiple regular bins. The spatial domain is represented by the coloured area in Figure \ref{fig:9} and it is divided in $0.1 \times 0.1$ longitude-latitude bins. The magnitude domain ranges from $4.95$ to $9.05$ magnitude units and is divided in bins of length $0.1$. The forecast is relative to the period from January 1, 2010, to December 31, 2014.

The adaptively-smoothed forecast provides the expected number of earthquakes in each space-magnitude bin. For each bin, to calculate the probability of observing at least one earthquake, in accordance with the methodology in the 2010 Italy CSEP forecast experiment, we consider a Poisson distribution for the number of events with intensity given by the predicted number of events. Assuming independence in the magnitude bins, we can aggregate the probabilities over magnitude bins and, for each space bin, obtain the probability of observing at least an earthquake in the period of interest with magnitude greater, or equal, to $4.95$. Figure \ref{fig:9} shows the forecasted log-probability for each spatial bin used as data generating model. 

% Figure 9

The Italian adaptively-smoothed forecast reported in Figure \ref{fig:9} is the vector of true probabilities $\mathbf p^* = p^*_1,...,p^*_N$, where $N = 8993$. As in the previous sections, we compare two forecasts $\mathbf p_1 = \mathbf p^*$ and $\mathbf p_2 = \omega \mathbf p^*$. We will be ignoring the spatial configuration. The average bin score difference is given by
$$
\begin{aligned}
\Delta(\mathbf p_1, \mathbf p_2, \mathbf X) & = \frac{1}{N}\sum_{i=1}^N (\Delta_{0,i} + X_i(\Delta_{1,i} - \Delta_{0,i})), \\ 
& = \bar \Delta_0 + \frac{1}{N}\sum_{i=1}^N X_i(\Delta_{1,i} - \Delta_{0,i}),
\end{aligned}
$$
where, $\Delta_{0,i} = \Delta(p_{1i}, p_{2i}, 0)$ and $\Delta_{1,i} = \Delta(p_{1i}, p_{2i}, 1)$ are, respectively, the score difference in the $i$-th bin in case we observe $X_i = 0$ (no earthquake at or above magnitude 4.95 during the 5 years) or $X_i = 1$ (at least one earthquake above magnitude 4.95 during the 5 years). The quantity $\bar \Delta_0$ is the average $\Delta_{0,i}$. The observations $X_i \sim \text{Ber}(p^*_i)$ follow a Bernoulli distribution, each bin has a potentially different parameter $p^*_i \neq p^*_j$ for any $i \neq j$. The expected value of the score difference is given by
$$
\mathbb E[\Delta(\mathbf p_1, \mathbf p_2, \mathbf X)] = \bar \Delta_0 + \frac{1}{N}\sum_{i=1}^N p^*_i(\Delta_{1,i} - \Delta_{0,i}).
$$

Given that we are considering $\mathbf p_1 = \mathbf p^*$ and $\mathbf p_2 = \omega \mathbf p^*$, the expected score difference is non-negative if a proper scoring rule is used while it could be negative if the scoring rule is improper. Specifically, we show that it is possible to find a reference model $\mathbf p_0$ such that, if used in combination with the parimutuel gambling score to rank the forecasts, the expected score difference is negative. As before, the Brier, log and full gambling score are used for comparison. In Table \ref{tab:3} we report the expected score differences considering different scores. As expected, they are all positive except for the pairwise gambling score.

% Table 3
\begin{table}
\caption{Expected score difference considering $\mathbf p^*$ equal to the 5 year Italy adaptively-smoothed forecast, $\mathbf p_1 = \mathbf p^*$, $\mathbf p_2 = \omega \mathbf p^*$ and reference model for the pairwise gambling score $\mathbf p_0 = 5\mathbf p^*$. The scores considered are: the Brier score, the log score, the pairwise gambling (PG) score and the full gambling (FG) score.}
\label{tab:3}       
% For LaTeX tables use
\begin{tabular}{l |   l}
\hline\noalign{\smallskip}
Score & $\mathbb E[\Delta]$  \\
\noalign{\smallskip}\hline\noalign{\smallskip}
Brier & 0.0000026\\
Log &  0.0003137\\
PG &  -0.0000900\\
FG &  0.0002422\\
\noalign{\smallskip}\hline
\end{tabular}
\end{table}

\begin{figure}
  \includegraphics[width = 0.99\textwidth]{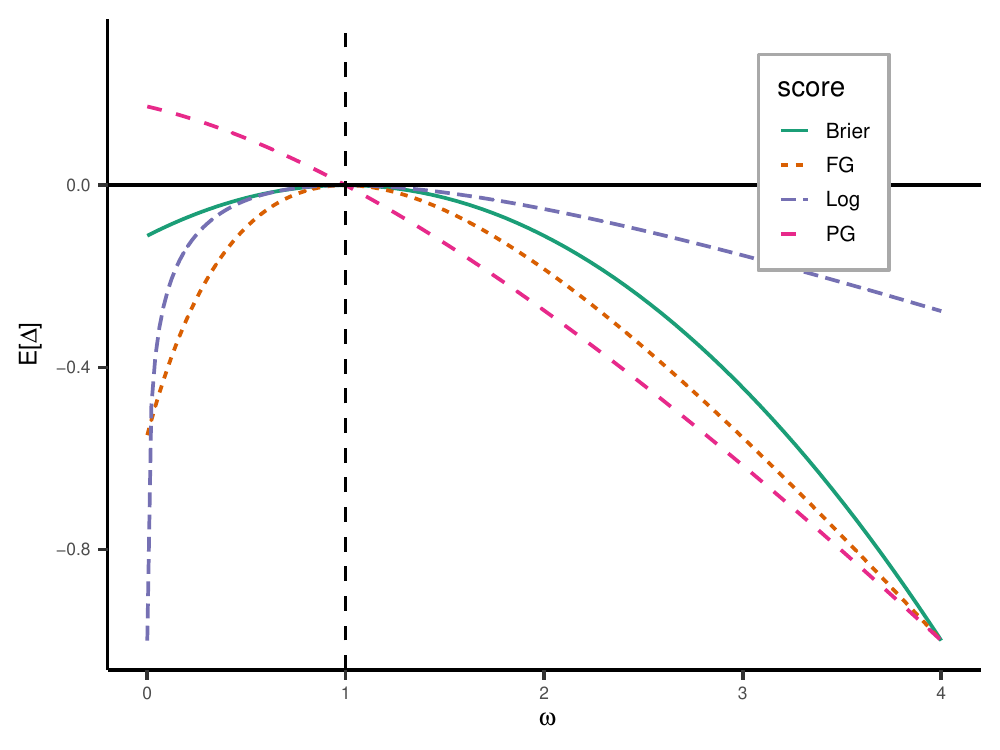}
\caption{Expected score difference between $\mathbf p_1$ and $\mathbf p_2$ as a function of $\omega = \mathbf p_1/\mathbf p_2$ for $\omega \in (10^{-3}, 4)$. We set $\mathbf p^*$ equal to the 5-year adaptively-smoothed Italy forecast, $\mathbf p_1 = \mathbf p^*$, $\mathbf p_2 = \omega \mathbf p^*$, and reference model for the pairwise gambling score $\mathbf p_0 = 5\mathbf p^*$.}
\label{fig:10}
\end{figure}

In Figure \ref{fig:10} is showed the expected score difference as a function of the forecasts ratio $\omega \in [10^{-3}, 4]$. The results are similar to the ones reported in Figure \ref{fig:1} and \ref{fig:2}. The Brier, log and full gambling scores behave suitably, while the pairwise gambling score does not. The Brier and full gambling score are bounded and they prefer a forecast $\mathbf p = 10^{-3}\mathbf p^*$ to $\mathbf p' = 4\mathbf p^*$. Indeed, in Figure \ref{fig:10} the left hand side is greater than the right hand side. That is because the penalty is based on the absolute difference between a forecast and the data generating model, therefore, a forecast $\mathbf p = 10^{-3}\mathbf p^*$ is preferred to $\mathbf p' = 4 \mathbf p^*$, because $\lVert 10^{-3}\mathbf p^* - \mathbf p^* \rVert \leq  \lVert 4\mathbf p^* - \mathbf p^* \rVert$. On the other hand, the log score is unbounded and is based on the relative difference. With the log score, a forecast $\mathbf p' = 4\mathbf p^*$ is preferred to $\mathbf p = 10^{-3}\mathbf p^*$ because $\lVert \mathbf p^*/ 10^{-3}\mathbf p^* \rVert > \lVert \mathbf p^*/ 4\mathbf p^*\rVert$. The pairwise gambling score, instead, is heavily biased towards zero.  

% Figure 10

\begin{figure}[!htbp]
  \includegraphics[width = 0.99\textwidth]{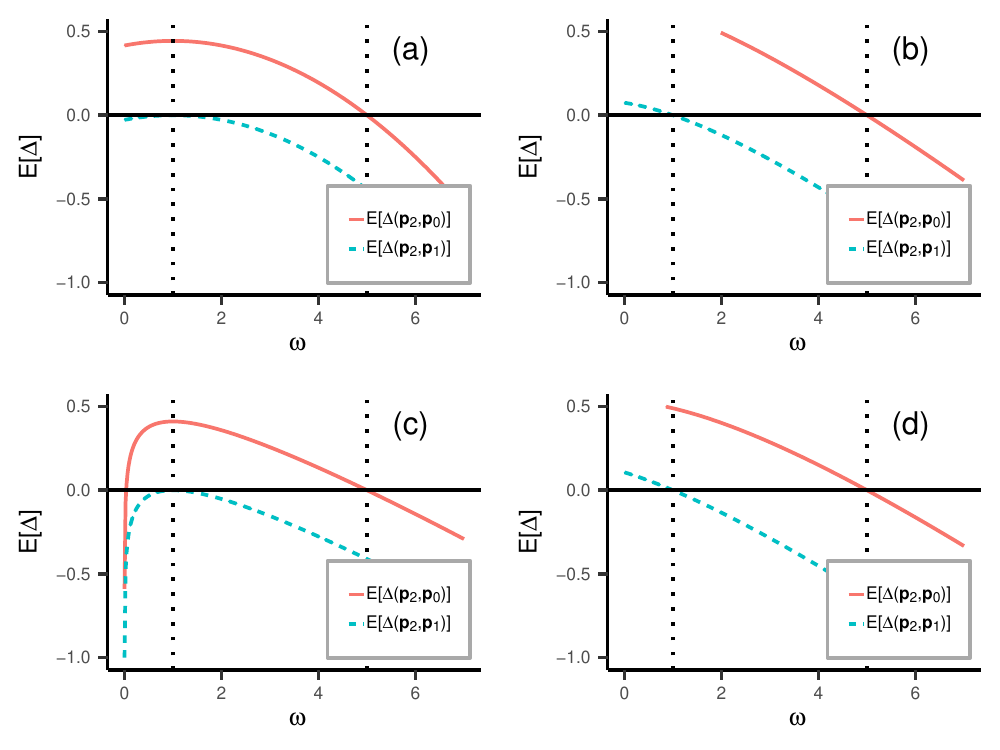}
\caption{Expected score difference between $\mathbf p_2$ and $\mathbf p_1$ (blue dashed) and $\mathbf p_2$ and $\mathbf p_0$ (red solid) as a function of $\omega = \mathbf p_1/\mathbf p_2$ for $\omega \in (10^{-3}, 7)$. We set $\mathbf p^*$ equal to 5-year adaptively-smoothed Italy forecast, $\mathbf p_1 = \mathbf p^*$, $\mathbf p_2 = \omega \mathbf p^*$ and $\mathbf p_0 = 5\mathbf p^*$. Vertical lines represent $\omega = 1$ and $\omega = 5$. We consider the Brier (a), pairwise gambling (b), log (c), and full gambling (d) scores.}
\label{fig:11}
\end{figure}

We can extend the comparison by considering $k = 3$ forecasts. In this case, we consider the reference model $\mathbf p_0 = 5\mathbf p^*$ as third competitor. Figure \ref{fig:11}, for each scoring rule, shows the expected score differences $\mathbb E[\Delta(\mathbf p_2, \mathbf p_1, \mathbf X)]$ (dashed blue) and $\mathbb E[\Delta(\mathbf p_2, \mathbf p_0, \mathbf X)]$ (solid red), representing the expected score difference between $\mathbf p_2$ and $\mathbf p_1$, and the expected score difference between $\mathbf p_2$ and $\mathbf p_0$. Given that $\mathbf p_1$ is equal to the true probabilities, the score differences have to be negative for any value of $\omega \neq 1$ in order for the scoring rule to be effective. Indeed, this is the case for the Brier and log score (Figure \ref{fig:11} (a), (c)). On the other hand, both the pairwise and full gambling score (Figure \ref{fig:11} (b), (d)) are improper and prefer $\mathbf p_2$ over $\mathbf p_1$ when $\omega \in (0,1)$. Moreover, all the scores prefer $\mathbf p_0$ to $\mathbf p_2$ when $\omega > 5$. However, the log score prefers $\mathbf p_0$ to $\mathbf p_2$ also when $\omega$ approaches zero. This shows, again, how different scoring rules apply different penalties to the forecasts.

% Figure 11

We note that the pairwise and full gambling score present almost the same expected score difference between $\mathbf p_2$ and $\mathbf p_1$. This is because both scoring procedures implicitly assume a reference model given by the average forecast. If the average forecast in a bin is greater than $p^*_i$, a forecaster will obtain a positive reward each time they submit a value smaller than $\mathbf p^*_i$ and $X_i = 0$ occurs. Therefore, given that we are in a low probability environment for which $\Pr[X_i = 0] > 0.99$, the smallest forecast is likely to be preferred. The bias depends on the relationship between the reference model and the true probabilities. In the gambling metaphore, the reference model plays the role of the house (or banker) which determines the returns, and against which all forecasts are competing. Considering equation \ref{eq:exp_parimutuel}, if $p^* < p_0$, the player has a positive reward forecasting $p < p_0$. If the number of forecasts is large enough that changing a forecast does not affect significantly the average, then, the smaller the forecast the higher the reward, and the forecaster is encouraged by the score to provide $p = 0$. The same reasoning applies if $p^* > p_0$. 

\subsection{ Confidence Interval and preference probabilities -  Multiple Bins Multiple Probabilities }

Also in the Multiple Bins Multiple Probabilities case it is crucial to account for the uncertainty around the observed score difference. The binomial formulation used before to retrieve confidence intervals no longer holds, and we need an alternative methodology. One approach to calculate confidence intervals for the expected score difference relies on a Gaussian approximation of the score difference distribution \citep{rhoades2011efficient}. The score difference in each bin, $\Delta(p_{1i}, p_{2i}, X_i)$ for $i = 1,...,N$, are assumed to be independent draws from a Gaussian distribution with expected value $\mathbb E[\Delta(\mathbf p_1, \mathbf p_2, X)]$ and variance $\sigma^2$. When we observe a sample $\mathbf x = x_1,...,x_N$, the point estimate of the expected score difference is the observed score difference $\Delta(\mathbf p_1, \mathbf p_2, \mathbf x)$ and the $(1-\alpha)\%$ confidence interval is given by:
$$
\Delta(\mathbf p_1, \mathbf p_2, \mathbf x) \pm t_{1-\alpha/2, N-1}\frac{s}{\sqrt N},
$$
where $s^2$ is an estimate of the variance $\sigma^2$ and $t_{1-\alpha/2, N-1}$ is the $1-\alpha/2$ percentile of a t-student distribution with $N - 1$ degrees of freedom. The reliability of such interval estimates is determined by the accuracy of the Gaussian approximation, which, in turns, depends on the amount of data (the more the better) and on the correlation between the score difference in each bin (the more the worst). We analyse the reliability of this approximation in Appendix A: Reliability of the gaussian confidence intervals and conclude that it can be used with the log, pairwise gambling and full gambling score but not with the Brier score. 

\begin{figure}
  \includegraphics[width = 0.99\textwidth]{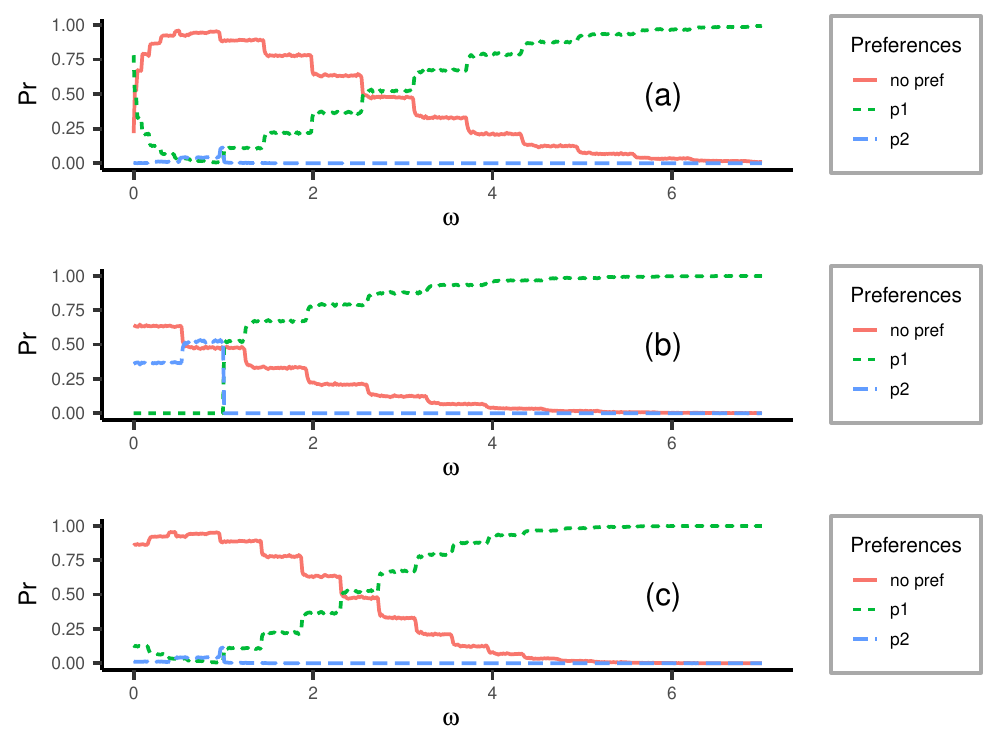}
\caption{Multiple Bins Multiple Probabilities case: each plot shows the probability of each possible outcome (solid no preference, dotted preference for $\mathbf p_1$ , dashed preference for $\mathbf p_2$ ) as a function of $\omega = \mathbf p_1/ \mathbf p_2$ for $\omega \in (10^{-3}, 7)$. The log (a), the pairwise gambling (b) and the full gambling (c) scores are considered. We set $\mathbf p^*$ equal to 5-year adaptively-smoothed Italy forecast, $\mathbf p_1 = \mathbf p^*$, $\mathbf p_2 = \omega \mathbf p^*$, and reference model for the pairwise gambling score $\mathbf p_0 = 5\mathbf p^*$.}
\label{fig:12}
\end{figure}

Figure \ref{fig:12} shows the evolution of the preference probabilities varying the forecasts ratio, $\omega = \mathbf p_2/\mathbf p_1$. It is quite similar to Figure \ref{fig:6} and the same problems with the pairwise gambling score are evident; i.e. it favours forecast smaller than the true probability when the average forecast is greater than the latter. On the other hand, the log score probability of preferring $\mathbf p_1$ increases rapidly when $\omega \rightarrow 0$, while the full gambling score is not able to distinguish between $\mathbf p_1$ and $\mathbf p_2$ for $\omega < 2.5$. The latter remark suggests a potential problem with the use of the full gambling score given that, in real forecasting experiments, the competing forecasts tends to be quite similar, in which case there is an high probability of no preference.  

This concludes our analysis on the use of proper scoring rules to rank earthquake forecasting models. 

% Figure 12

\section{ Discussion }

The parimutuel gambling score was introduced as a general scoring rule to compare, within a unified framework, earthquake forecasts of different kinds (e.g. alarm-based forecast and probabilistic forecast). It overcomes two limitations common to other forecast comparison techniques: i) the need to define a reference model, and  ii) to allow forecasts defined on different space-time-magnitude regions to be compared. We showed that the parimutuel gambling score is proper only when two forecasts are compared directly against each other. In the other cases (multi-forecast comparison and comparison against a reference model), the parimutuel gambling score is improper. Consequently, we discourage its use in multi-model comparisons such as CSEP and encourage researchers and practitioners to re-consider rankings obtained using this score.

Specifically, the parimutuel gambling score tries to avoid the need to pre-define a reference model by using the average forecast. Therefore, for each bin, a positive reward means that the model is \emph{better} than the average forecast, vice versa if the reward is negative. This allows to produce a map of the parimutuel gambling rewards from which to infer the bins where the forecast is better than the average forecast, and the bins where it is not. Since the parimutuel gambling score is proper only when $k = 2$, any map obtained by computing the comparisons for $k > 2$ may be biased. This difficulties may be circumvented by using any proper scoring rule that allows for multi-forecast comparison. In fact, maps of this kind may be produced by reporting the score difference between a forecast and the average one. Furthermore, given that proper scores are scale invariant, we can re-scale the score values to be between $-1$ and $1$. In this way, we can visualize which bins has a positive or negative contribution to the average score difference. 

The need to compare forecasts defined on different set of bins comes from the design of the forecasting experiment. In the RELM experiment \citep{zechar2013regional} modelers were allowed to choose a subset of bins to include in their forecast, referred to as masking. Modellers involved in the RELM experiment provided forecasts with very different masks; some issued forecasts for the entire California region \citep{bird2007seismic, helmstetter2007high, holliday2007relm}, some for only Southern California \citep{ward2007methods, shen2007implications, kagan2007testable}, while others used irregular masks \citep{ebel2007non}. The parimutuel gambling score addressed these differences using the gambling metaphor. Each forecaster is a gambler which plays a certain number of rounds (bets) corresponding to the bins. A forecaster does not have to make a forecast for every bin - they can just sit out this round. The forecasters are ranked by their total reward (i.e. the sum of the rewards for each bin). We argue that this solution is still problematic. First, the parimutuel gambling score needs at least two forecasts for each bin to be computed. If only one forecaster \emph{plays} in a bin we can not calculate the parimutuel gambling score for that bin. Second, consider two bins for which different sets of forecasters provided a forecast; in this situation the models are rewarded with respect to different odds. This becomes problematic when we attempt to interpret the observed result because in each bin the reference model is given by a potentially different combination of models.

The Brier score can also be used to assess masked forecasts. The maximum Brier score value obtainable by a forecast is zero and it is achieved by the \emph{perfect} forecast which assumes $p = 1$ when $x = 1$ and $p = 0$ when $x = 0$. Any other forecast obtains a negative Brier score. Therefore, the Brier score of a forecast can be seen as the Brier score difference between the perfect forecast and the forecast under evaluation. Given two models, the one with the highest average Brier score is the one \emph{closest} (on average) to the perfect forecast. If the models provide forecasts on two different sets of bins, $B_1$ and $B_2$, we can still compare the forecasts in terms of their average Brier score. Suppose the first model achieves an higher average Brier score, we can conclude that, on average, the first model in $B_1$ is closer to the perfect forecast than the second model in $B_2$. We can make the above comparison also when $B_1$ and $B_2$ have zero bins in common because the Brier score requires only a forecast and the observation to be calculated. 

In this paper, we have used confidence intervals to asses the statistical significance of observed score differences. Analytically determining these confidence intervals may be too complex and a basic approximative Gaussian approach may fail, as highlighted by the Brier score example. Problems of this type come from the fact that the score differences per bin are treated as independent and identically distributed. This assumption is false, especially when considering space-time bins which depend on each other both in time and space due to clustering of earthquakes. A possible solution to relax the independence assumption is to consider a Diebold-Mariano test \citep{diebold2002comparing} on the score differences which takes into account the correlation structure of the score differences sequence.

\section{ Conclusion }

The parimutuel gambling score, commonly applied to compare earthquake forecasts, is improper when the number of forecasts being tested is greater than two. In the special case of two competing forecasts, the score is proper, and can return results similar to alternate proper scoring methods, but even then it can be used improperly. In the common testing scenario of multiple forecasts being compared simultaneously, or when multiple forecasts are compared against a reference model, the parimutuel gambling score provides a biased assessment of the skill of a forecast when it is tested against a given outcome. This is fundamentally a problem of the gambling analogy itself; the betting strategy of maximizing the expected reward (score) does not have to be consistent with the data generating model (in the case where this is known) and, therefore, gamblers (modellers) are not encouraged to provide forecasts resembling the data generating model. This is because the score for a given forecast is dependent on all the forecasts taking part in the competition, not just on the observed data; one can therefore change the ranking of two models by changing one of the other models in the pool. This introduces the undesirable property that one can potentially game the system to prefer a specific model. Further, if we only have access to the forecasts and the data, it is impossible to know if the parimutuel gambling score results will be biased or not. Moreover, the only case in which they are correct is when one of the competing forecasts is the data generating model, which is highly unlikely. These findings are sufficiently clear for us to discourage the use of the parimutuel gambling score in distinguishing between multiple competing forecasts, and for care to be taken even in the case where only two are being compared.

We recommend that alternative scores that do not suffer from these shortcomings should be used instead to assess the skill of prospective earthquake forecasts in a formal testing environment. The Brier and log scores are both proper, and require no new information beyond what was used to calculate the parimutuel gambling score, so switching existing analyses to a proper score should be simple to implement. We recommend testing for properness when introducing new scoring rules, either analytically or via simulations using a known model to generate testing data. 

\section{Acknowledgments}
All the code to produce the present results is written in the R programming language. We have used the package $\texttt{ggplot2}$ \citep{ggplot}, the package $\texttt{rnaturalearth}$ \citep{south2017rnaturalearth} for the Italy contour map presented in Figure \ref{fig:9} and the package $\texttt{bayesianETAS}$ \citep{ross2016bayesian}
for the Maximum Likelihood estimate used in Section 5. We thank Kirsty Bayliss and for her useful feedback and constructive discussions which led to the final version of this work.

\section*{Declarations}

Some journals require declarations to be submitted in a standardised format. Please check the Instructions for Authors of the journal to which you are submitting to see if you need to complete this section. If yes, your manuscript must contain the following sections under the heading `Declarations':

\begin{itemize}
\item Funding: This work was founded by the Real-time Earthquake Risk Reduction for a Resilient Europe “RISE” project, which has received funding from the European Union's Horizon 2020 research and innovation program under grant Agreement 821115.
\item Conflict of interest/Competing interests : The authors have no financial or proprietary interests in any material discussed in this article.
\item Ethics Approval
\item Consent to participate : Not applicable
\item Consent to publication : Not applicable
\item Availability of data and materials : Simulated data and figures can be found at \url{https://github.com/Serra314/Serra314.github.io/tree/master/Ranking_earthquake_forecast}.
\item Code availability : The code to generate all the figures in the manuscript can be found at 
\url{https://github.com/Serra314/Serra314.github.io/tree/master/Ranking_earthquake_forecast}.
\item Authors' contributions : All authors contributed to the study design. Material preparation, analysis and first draft were performed by Francesco Serafini. Maximilian Werner provided the 5-year adaptively-smoothed Italy forecast. All authors commented and contributed to previous versions of the manuscript. All authors read and approved the submitted manuscript.  
\end{itemize}

\section{Appendix A: Reliability of the gaussian confidence intervals}\label{secA1}

We assess the reliability of the Gaussian approximation to calculate confidence intervals for the expected score differences using simulated replicates from $\mathbf p^*$ equal to the 5-year adaptively-smoothed Italy forecast. We set the first forecasts $\mathbf p_1 = \mathbf p^*$, $\mathbf p_2 = \omega \mathbf p^*$, and the reference model $\mathbf p_0 = 5 \mathbf p^*$.
The method is reliable if the probability that the approximated confidence interval of level $\alpha$ contains the true expected value (coverage probability) is close to $1 - \alpha$. 
To estimate the coverage probability, for a set of values of $\omega \in (10^{-3}, 7)$, we simulated $\mathbf x = x_1,...,x_N$ 10,000 times and calculated the approximate 95\% confidence intervals. The coverage probability is given by the fraction of times in which the confidence intervals contains the true expected score difference. 

% Figure 13
\begin{figure}[H]
  \includegraphics[width = 0.99\textwidth]{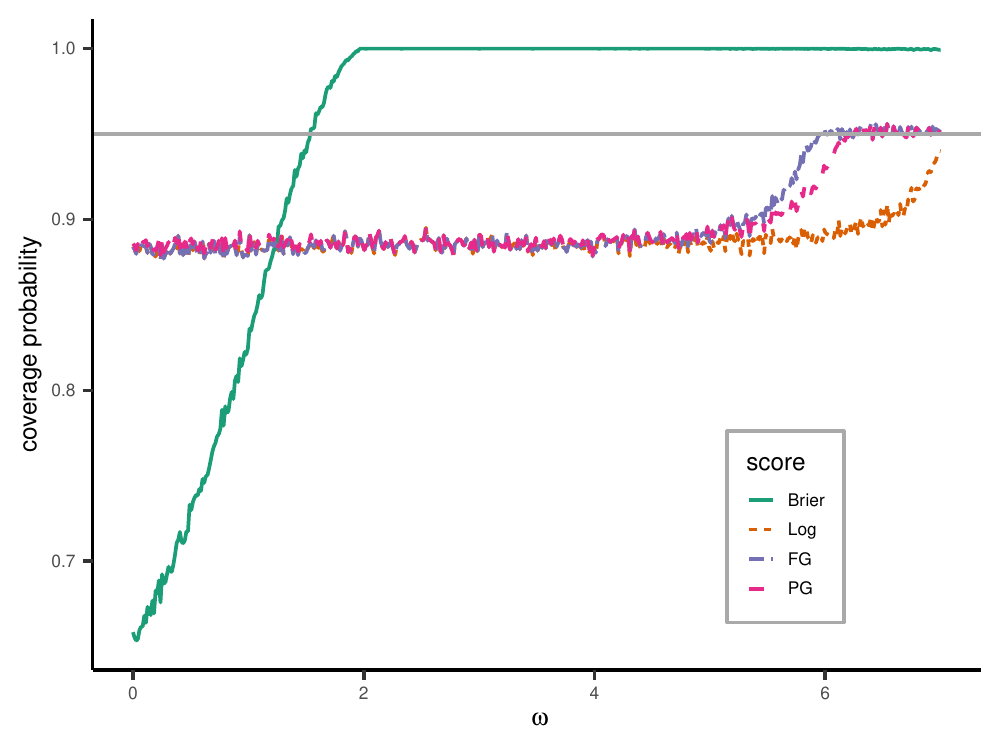}
\caption{Coverage probability for the Gaussian approximated confidence interval for the expected score difference  between $\mathbf p_1$ and $\mathbf p_2$ as a function of $\omega = \mathbf p_2/ \mathbf p_1$ for $\omega \in (10^{-3}, 7)$. We consider the Brier, log, pairwise gambling (PG) and full gambling (FG) scores. We set $\mathbf p^*$ equal to the 5-year adaptively-smoothed Italy forecast, $\mathbf p_1 = \mathbf p^*$, $\mathbf p_2 = \omega \mathbf p^*$ and $\mathbf p_0 = 5\mathbf p^*$. Horizontal line represents the target coverage probability ($0.95$) }
\label{fig:13}
\end{figure}

\begin{figure}[H]
  \includegraphics[width = 0.99\textwidth]{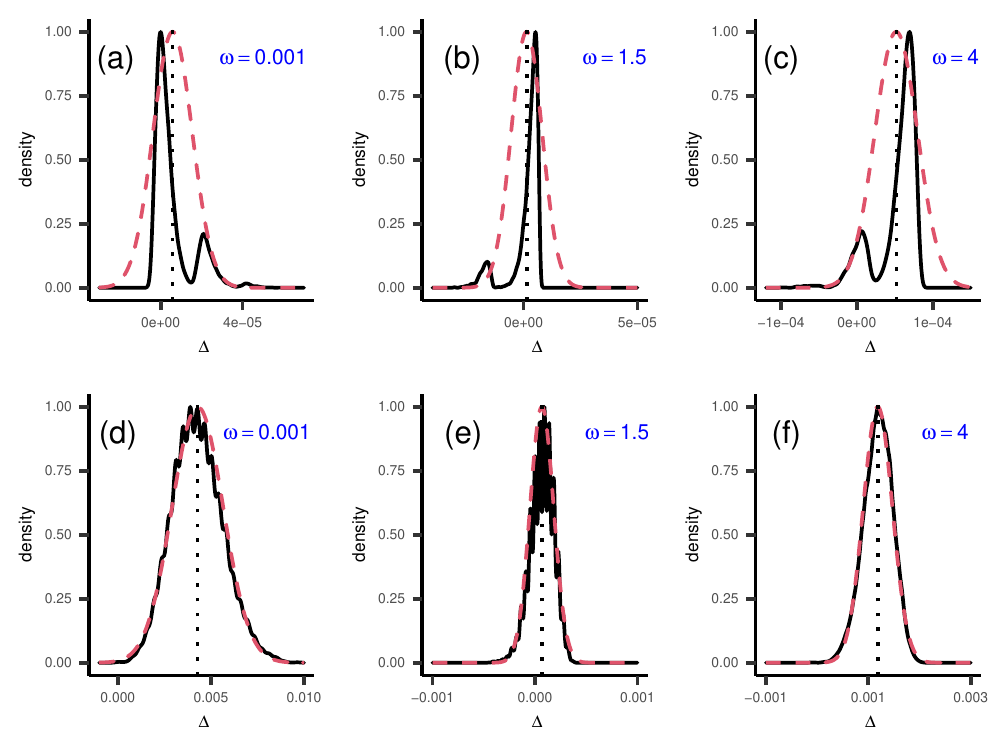}
\caption{Average score difference distribution for the Brier score (a-c) and log score (d-f). We set $\mathbf p^*$ equal to the 5-year adaptively-smoothed Italy forecast, $\mathbf p_1 = \mathbf p^*$, $\mathbf p_2 = \omega \mathbf p^*$, and reference model for the pairwise gambling score $\mathbf p_0 = 5\mathbf p^*$. We consider $\omega = 0.001$ (a,d); $\omega = 1.5$ (b,e); $\omega = 4$ (c,f). Black solid line represents the empirical distribution obtained from 10000 simulations. Red dashed line represents the corresponding Gaussian approximation. Vertical dotted line represents the true value of the expected score difference.}
\label{fig:14}
\end{figure}

Figure \ref{fig:13} shows that the approximation is not reliable for the Brier score for which produces confidence intervals which are too small (coverage below $0.95$) or too wide (coverage above $0.95$) depending on the value of $\omega$. This is because the Brier score differences have a asymmetrical bimodal distribution as shown in Figure \ref{fig:14}a-c and therefore are not normally distributed. Furthermore, Figure \ref{fig:13} shows also that the coverage for the log, pairwise gambling and full gambling score is usually below 0.95 and reaches this value only for $\omega > 6$. This is because, taking the log score as example, the distribution of the score differences becomes smoother as $\omega$ grows (Figure \ref{fig:14}d-f). The distributions of the pairwise and full gambling score resemble the log score one. 

From this analysis we conclude that it is possible to use the gaussianly approximated confidence intervals for the log, pairwise gambling and full gambling because the coverage probability is always between $0.88$ and $0.96$. In this example, the reliability of the approximation depends on how much different the forecasts are. 

% Figure 14

\bibliographystyle{unsrtnat}

\begin{thebibliography}{64}

\bibitem[Rhoades et~al.(2016)Rhoades, Liukis, Christophersen, and
  Gerstenberger]{rhoades2016retrospective}
DA~Rhoades, M~Liukis, A~Christophersen, and MC~Gerstenberger.
\newblock Retrospective tests of hybrid operational earthquake forecasting
  models for canterbury.
\newblock \emph{Geophysical Journal International}, 204\penalty0 (1):\penalty0
  440--456, 2016.

\bibitem[Bourne et~al.(2018)Bourne, Oates, and Van~Elk]{bourne2018exponential}
SJ~Bourne, SJ~Oates, and J~Van~Elk.
\newblock The exponential rise of induced seismicity with increasing stress
  levels in the groningen gas field and its implications for controlling
  seismic risk.
\newblock \emph{Geophysical Journal International}, 213\penalty0 (3):\penalty0
  1693--1700, 2018.

\bibitem[Marzocchi et~al.(2014)Marzocchi, Lombardi, and
  Casarotti]{marzocchi2014establishment}
Warner Marzocchi, Anna~Maria Lombardi, and Emanuele Casarotti.
\newblock The establishment of an operational earthquake forecasting system in
  italy.
\newblock \emph{Seismological Research Letters}, 85\penalty0 (5):\penalty0
  961--969, 2014.

\bibitem[Iervolino et~al.(2015)Iervolino, Chioccarelli, Giorgio, Marzocchi,
  Zuccaro, Dolce, and Manfredi]{iervolino2015operational}
Iunio Iervolino, Eugenio Chioccarelli, Massimiliano Giorgio, Warner Marzocchi,
  Giulio Zuccaro, Mauro Dolce, and Gaetano Manfredi.
\newblock Operational (short-term) earthquake loss forecasting in italy.
\newblock \emph{Bulletin of the Seismological Society of America}, 105\penalty0
  (4):\penalty0 2286--2298, 2015.

\bibitem[Jordan(2006)]{jordan2006brick}
Thomas~H Jordan.
\newblock Earthquake predictability, brick by brick.
\newblock \emph{Seismological Research Letters}, 77\penalty0 (1):\penalty0
  3--6, 2006.

\bibitem[Zechar et~al.(2010{\natexlab{a}})Zechar, Schorlemmer, Liukis, Yu,
  Euchner, Maechling, and Jordan]{zechar2010csep}
J~Douglas Zechar, Danijel Schorlemmer, Maria Liukis, John Yu, Fabian Euchner,
  Philip~J Maechling, and Thomas~H Jordan.
\newblock The collaboratory for the study of earthquake predictability
  perspective on computational earthquake science.
\newblock \emph{Concurrency and Computation: Practice and Experience},
  22\penalty0 (12):\penalty0 1836--1847, 2010{\natexlab{a}}.

\bibitem[Schorlemmer et~al.(2018)Schorlemmer, Werner, Marzocchi, Jordan, Ogata,
  Jackson, Mak, Rhoades, Gerstenberger, Hirata, et~al.]{achievements}
Danijel Schorlemmer, Maximilian~J Werner, Warner Marzocchi, Thomas~H Jordan,
  Yosihiko Ogata, David~D Jackson, Sum Mak, David~A Rhoades, Matthew~C
  Gerstenberger, Naoshi Hirata, et~al.
\newblock The collaboratory for the study of earthquake predictability:
  achievements and priorities.
\newblock \emph{Seismological Research Letters}, 89\penalty0 (4):\penalty0
  1305--1313, 2018.

\bibitem[Gneiting and Raftery(2007)]{gneiting2007strictly}
Tilmann Gneiting and Adrian~E Raftery.
\newblock Strictly proper scoring rules, prediction, and estimation.
\newblock \emph{Journal of the American statistical Association}, 102\penalty0
  (477):\penalty0 359--378, 2007.

\bibitem[Murphy(1993)]{murphy1993good}
Allan~H Murphy.
\newblock What is a good forecast? an essay on the nature of goodness in
  weather forecasting.
\newblock \emph{Weather and forecasting}, 8\penalty0 (2):\penalty0 281--293,
  1993.

\bibitem[Jolliffe and Stephenson(2003)]{jolliffe2003forecast}
IT~Jolliffe and DB~Stephenson.
\newblock Forecast verification. chichester, england and hoboken, 2003.

\bibitem[Rosen(1996)]{rosen1996good}
David~B Rosen.
\newblock How good were those probability predictions? the expected
  recommendation loss (erl) scoring rule.
\newblock In \emph{Maximum Entropy and Bayesian Methods}, pages 401--408.
  Springer, 1996.

\bibitem[Hyv{\"a}rinen and Dayan(2005)]{hyvarinen2005estimation}
Aapo Hyv{\"a}rinen and Peter Dayan.
\newblock Estimation of non-normalized statistical models by score matching.
\newblock \emph{Journal of Machine Learning Research}, 6\penalty0 (4), 2005.

\bibitem[Hern{\'a}ndez-Orallo et~al.(2012)Hern{\'a}ndez-Orallo, Flach, and
  Ferri~Ram{\'\i}rez]{hernandez2012unified}
Jos{\'e} Hern{\'a}ndez-Orallo, Peter Flach, and C{\'e}sar Ferri~Ram{\'\i}rez.
\newblock A unified view of performance metrics: Translating threshold choice
  into expected classification loss.
\newblock \emph{Journal of Machine Learning Research}, 13:\penalty0 2813--2869,
  2012.

\bibitem[Field et~al.(2014)Field, Arrowsmith, Biasi, Bird, Dawson, Felzer,
  Jackson, Johnson, Jordan, Madden, et~al.]{field2014uniform}
Edward~H Field, Ramon~J Arrowsmith, Glenn~P Biasi, Peter Bird, Timothy~E
  Dawson, Karen~R Felzer, David~D Jackson, Kaj~M Johnson, Thomas~H Jordan,
  Christopher Madden, et~al.
\newblock Uniform california earthquake rupture forecast, version 3
  (ucerf3)—the time-independent model.
\newblock \emph{Bulletin of the Seismological Society of America}, 104\penalty0
  (3):\penalty0 1122--1180, 2014.

\bibitem[Steacy et~al.(2014)Steacy, Gerstenberger, Williams, Rhoades, and
  Christophersen]{steacy2014new}
Sandy Steacy, Matt Gerstenberger, Charles Williams, David Rhoades, and
  Annemarie Christophersen.
\newblock A new hybrid coulomb/statistical model for forecasting aftershock
  rates.
\newblock \emph{Geophysical Journal International}, 196\penalty0 (2):\penalty0
  918--923, 2014.

\bibitem[Bayliss et~al.(2020)Bayliss, Naylor, Illian, and
  Main]{bayliss2020data}
Kirsty Bayliss, Mark Naylor, Janine Illian, and Ian~G Main.
\newblock Data-driven optimization of seismicity models using diverse data
  sets: Generation, evaluation, and ranking using inlabru.
\newblock \emph{Journal of Geophysical Research: Solid Earth}, 125\penalty0
  (11):\penalty0 e2020JB020226, 2020.

\bibitem[Schorlemmer and Gerstenberger(2007{\natexlab{a}})]{schor2007relm}
Danijel Schorlemmer and MC~Gerstenberger.
\newblock Relm testing center.
\newblock \emph{Seismological Research Letters}, 78\penalty0 (1):\penalty0
  30--36, 2007{\natexlab{a}}.

\bibitem[Savran et~al.(2020)Savran, Werner, Marzocchi, Rhoades, Jackson,
  Milner, Field, and Michael]{savran2020pseudoprospective}
William~H Savran, Maximilian~J Werner, Warner Marzocchi, David~A Rhoades,
  David~D Jackson, Kevin Milner, Edward Field, and Andrew Michael.
\newblock Pseudoprospective evaluation of ucerf3-etas forecasts during the 2019
  ridgecrest sequence.
\newblock \emph{Bulletin of the Seismological Society of America}, 110\penalty0
  (4):\penalty0 1799--1817, 2020.

\bibitem[Zechar and Jordan(2008)]{zechar2008testing}
J~Douglas Zechar and Thomas~H Jordan.
\newblock Testing alarm-based earthquake predictions.
\newblock \emph{Geophysical Journal International}, 172\penalty0 (2):\penalty0
  715--724, 2008.

\bibitem[Zechar and Jordan(2010)]{zechar2010area}
J~Douglas Zechar and Thomas~H Jordan.
\newblock The area skill score statistic for evaluating earthquake
  predictability experiments.
\newblock In \emph{Seismogenesis and Earthquake Forecasting: The Frank Evison
  Volume II}, pages 39--52. Springer, 2010.

\bibitem[Stark(1997)]{stark1997earthquake}
Philip~B Stark.
\newblock Earthquake prediction: the null hypothesis.
\newblock \emph{Geophysical Journal International}, 131\penalty0 (3):\penalty0
  495--499, 1997.

\bibitem[Luen et~al.(2008)Luen, Stark, et~al.]{luen2008testing}
Brad Luen, Philip~B Stark, et~al.
\newblock Testing earthquake predictions.
\newblock In \emph{Probability and Statistics: Essays in Honor of David A.
  Freedman}, pages 302--315. Institute of Mathematical Statistics, 2008.

\bibitem[Marzocchi and Zechar(2011)]{marzocchi2011earthquake}
Warner Marzocchi and J~Douglas Zechar.
\newblock Earthquake forecasting and earthquake prediction: different
  approaches for obtaining the best model.
\newblock \emph{Seismological Research Letters}, 82\penalty0 (3):\penalty0
  442--448, 2011.

\bibitem[Schorlemmer et~al.(2007)Schorlemmer, Gerstenberger, Wiemer, Jackson,
  and Rhoades]{sch}
D~Schorlemmer, MC~Gerstenberger, S~Wiemer, DD~Jackson, and DA~Rhoades.
\newblock Earthquake likelihood model testing.
\newblock \emph{Seismological Research Letters}, 78\penalty0 (1):\penalty0
  17--29, 2007.

\bibitem[Zechar et~al.(2010{\natexlab{b}})Zechar, Gerstenberger, and
  Rhoades]{zechar}
J~Douglas Zechar, Matthew~C Gerstenberger, and David~A Rhoades.
\newblock Likelihood-based tests for evaluating space--rate--magnitude
  earthquake forecasts.
\newblock \emph{Bulletin of the Seismological Society of America}, 100\penalty0
  (3):\penalty0 1184--1195, 2010{\natexlab{b}}.

\bibitem[Rhoades et~al.(2011)Rhoades, Schorlemmer, Gerstenberger,
  Christophersen, Zechar, and Imoto]{rhoades2011efficient}
David~A Rhoades, Danijel Schorlemmer, Matthew~C Gerstenberger, Annemarie
  Christophersen, J~Douglas Zechar, and Masajiro Imoto.
\newblock Efficient testing of earthquake forecasting models.
\newblock \emph{Acta Geophysica}, 59\penalty0 (4):\penalty0 728--747, 2011.

\bibitem[Schneider et~al.(2014)Schneider, Clements, Rhoades, and
  Schorlemmer]{schneider2014likelihood}
Max Schneider, Robert Clements, David Rhoades, and Danijel Schorlemmer.
\newblock Likelihood-and residual-based evaluation of medium-term earthquake
  forecast models for california.
\newblock \emph{Geophysical Journal International}, 198\penalty0 (3):\penalty0
  1307--1318, 2014.

\bibitem[Werner and Sornette(2008)]{werner2008magnitude}
Maximilian~J Werner and Didier Sornette.
\newblock Magnitude uncertainties impact seismic rate estimates, forecasts, and
  predictability experiments.
\newblock \emph{Journal of Geophysical Research: Solid Earth}, 113\penalty0
  (B8), 2008.

\bibitem[Zechar et~al.(2013)Zechar, Schorlemmer, Werner, Gerstenberger,
  Rhoades, and Jordan]{zechar2013regional}
J~Douglas Zechar, Danijel Schorlemmer, Maximilian~J Werner, Matthew~C
  Gerstenberger, David~A Rhoades, and Thomas~H Jordan.
\newblock Regional earthquake likelihood models i: First-order results.
\newblock \emph{Bulletin of the Seismological Society of America}, 103\penalty0
  (2A):\penalty0 787--798, 2013.

\bibitem[Marzocchi et~al.(2012)Marzocchi, Zechar, and
  Jordan]{marzocchi2012bayesian}
Warner Marzocchi, J~Douglas Zechar, and Thomas~H Jordan.
\newblock Bayesian forecast evaluation and ensemble earthquake forecasting.
\newblock \emph{Bulletin of the Seismological Society of America}, 102\penalty0
  (6):\penalty0 2574--2584, 2012.

\bibitem[Holliday et~al.(2005)Holliday, Nanjo, Tiampo, Rundle, and
  Turcotte]{holliday2005earthquake}
James~R Holliday, Kazuyoshi~Z Nanjo, Kristy~F Tiampo, John~B Rundle, and
  Donald~L Turcotte.
\newblock Earthquake forecasting and its verification.
\newblock \emph{arXiv preprint cond-mat/0508476}, 2005.

\bibitem[Zechar and Zhuang(2014)]{zechar2014}
J~Douglas Zechar and Jiancang Zhuang.
\newblock A parimutuel gambling perspective to compare probabilistic seismicity
  forecasts.
\newblock \emph{Geophysical Journal International}, 199\penalty0 (1):\penalty0
  60--68, 2014.

\bibitem[Field(2007)]{field2007overview}
Edward~H Field.
\newblock Overview of the working group for the development of regional
  earthquake likelihood models (relm).
\newblock \emph{Seismological Research Letters}, 78\penalty0 (1):\penalty0
  7--16, 2007.

\bibitem[Zhuang(2010)]{zhuang2010}
Jiancang Zhuang.
\newblock Gambling scores for earthquake predictions and forecasts.
\newblock \emph{Geophysical Journal International}, 181\penalty0 (1):\penalty0
  382--390, 2010.

\bibitem[Main(1997)]{main1997long}
Ian Main.
\newblock Long odds on prediction.
\newblock \emph{Nature}, 385\penalty0 (6611):\penalty0 19--20, 1997.

\bibitem[Kossobokov(2004)]{kossobokov2004earthquake}
Kossobokov.
\newblock Earthquake prediction: basics, achievements, perspectives.
\newblock \emph{Acta Geodaetica et Geophysica Hungarica}, 39\penalty0
  (2-3):\penalty0 205--221, 2004.

\bibitem[{Kossobokov}(2006)]{kossobokov2006testing}
{Kossobokov}.
\newblock Testing earthquake prediction methods: The west pacific short-term
  forecast of earthquakes with magnitude mwhr $\geq$ 5.8.
\newblock \emph{Tectonophysics}, 413\penalty0 (1-2):\penalty0 25--31, 2006.

\bibitem[Brier(1950)]{brier1950verification}
Glenn~W Brier.
\newblock Verification of forecasts expressed in terms of probability.
\newblock \emph{Monthly weather review}, 78\penalty0 (1):\penalty0 1--3, 1950.

\bibitem[Good(1952)]{goodj}
I.~J. Good.
\newblock Rational decisions.
\newblock \emph{Journal of the Royal Statistical Society, Ser. B}, pages
  107--114, 1952.

\bibitem[Werner et~al.(2010)Werner, Helmstetter, Jackson, Kagan, and
  Wiemer]{werner2010adaptively}
Maximilian~J Werner, Agn{\`e}s Helmstetter, David~D Jackson, Yan~Y Kagan, and
  Stefan Wiemer.
\newblock Adaptively smoothed seismicity earthquake forecasts for italy.
\newblock \emph{arXiv preprint arXiv:1003.4374}, 2010.

\bibitem[Huber(1992)]{huber1992robust}
Peter~J Huber.
\newblock Robust estimation of a location parameter.
\newblock In \emph{Breakthroughs in statistics}, pages 492--518. Springer,
  1992.

\bibitem[Marzocchi and Lombardi(2009)]{marzocchi2009real}
Warner Marzocchi and Anna~Maria Lombardi.
\newblock Real-time forecasting following a damaging earthquake.
\newblock \emph{Geophysical Research Letters}, 36\penalty0 (21), 2009.

\bibitem[Taroni et~al.(2018)Taroni, Marzocchi, Schorlemmer, Werner, Wiemer,
  Zechar, Heiniger, and Euchner]{taroni2018prospective}
Matteo Taroni, Warner Marzocchi, Danijel Schorlemmer, Maximilian~Jonas Werner,
  Stefan Wiemer, Jeremy~Douglas Zechar, Lukas Heiniger, and Fabian Euchner.
\newblock Prospective csep evaluation of 1-day, 3-month, and 5-yr earthquake
  forecasts for italy.
\newblock \emph{Seismological Research Letters}, 89\penalty0 (4):\penalty0
  1251--1261, 2018.

\bibitem[Zechar and Zhuang(2010)]{zechar2010risk}
J~Douglas Zechar and Jiancang Zhuang.
\newblock Risk and return: evaluating reverse tracing of precursors earthquake
  predictions.
\newblock \emph{Geophysical Journal International}, 182\penalty0 (3):\penalty0
  1319--1326, 2010.

\bibitem[Taroni et~al.(2014)Taroni, Zechar, and Marzocchi]{taroni2014assessing}
M~Taroni, JD~Zechar, and W~Marzocchi.
\newblock Assessing annual global m 6+ seismicity forecasts.
\newblock \emph{Geophysical Journal International}, 196\penalty0 (1):\penalty0
  422--431, 2014.

\bibitem[Gruppo~di Lavoro(2004)]{gruppo2004redazione}
MPS Gruppo~di Lavoro.
\newblock Redazione della mappa di pericolosit{\`a} sismica prevista
  dall’ordinanza pcm 3274 del 20 marzo 2003.
\newblock \emph{Rapporto Conclusivo per il Dipartimento della Protezione
  Civile, INGV, Milano-Roma}, 5, 2004.

\bibitem[Fisher(1922)]{fisher1922}
Ronald~A Fisher.
\newblock On the mathematical foundations of theoretical statistics.
\newblock \emph{Philosophical Transactions of the Royal Society of London.
  Series A, Containing Papers of a Mathematical or Physical Character},
  222\penalty0 (594-604):\penalty0 309--368, 1922.

\bibitem[Schervish(2012)]{schervish2012}
Mark~J Schervish.
\newblock \emph{Theory of statistics}.
\newblock Springer Science \& Business Media, 2012.

\bibitem[Hastie et~al.(2009)Hastie, Tibshirani, and
  Friedman]{hastie2009elements}
Trevor Hastie, Robert Tibshirani, and Jerome Friedman.
\newblock \emph{The elements of statistical learning: data mining, inference,
  and prediction}.
\newblock Springer Science \& Business Media, 2009.

\bibitem[Wallis(2013)]{wallis2013binomial}
Sean Wallis.
\newblock Binomial confidence intervals and contingency tests: mathematical
  fundamentals and the evaluation of alternative methods.
\newblock \emph{Journal of Quantitative Linguistics}, 20\penalty0 (3):\penalty0
  178--208, 2013.

\bibitem[Clopper and Pearson(1934)]{clopper1934use}
Charles~J Clopper and Egon~S Pearson.
\newblock The use of confidence or fiducial limits illustrated in the case of
  the binomial.
\newblock \emph{Biometrika}, 26\penalty0 (4):\penalty0 404--413, 1934.

\bibitem[Schorlemmer and
  Gerstenberger(2007{\natexlab{b}})]{schorlemmer2007relm}
Danijel Schorlemmer and MC~Gerstenberger.
\newblock Relm testing center.
\newblock \emph{Seismological Research Letters}, 78\penalty0 (1):\penalty0
  30--36, 2007{\natexlab{b}}.

\bibitem[Michael and Werner(2018)]{michael2018preface}
Andrew~J Michael and Maximilian~J Werner.
\newblock Preface to the focus section on the collaboratory for the study of
  earthquake predictability (csep): New results and future directions.
\newblock \emph{Seismological Research Letters}, 89\penalty0 (4):\penalty0
  1226--1228, 2018.

\bibitem[Bird and Liu(2007)]{bird2007seismic}
Peter Bird and Zhen Liu.
\newblock Seismic hazard inferred from tectonics: California.
\newblock \emph{Seismological Research Letters}, 78\penalty0 (1):\penalty0
  37--48, 2007.

\bibitem[Helmstetter et~al.(2007)Helmstetter, Kagan, and
  Jackson]{helmstetter2007high}
Agnes Helmstetter, Yan~Y Kagan, and David~D Jackson.
\newblock High-resolution time-independent grid-based forecast for m $\geq$ 5
  earthquakes in california.
\newblock \emph{Seismological Research Letters}, 78\penalty0 (1):\penalty0
  78--86, 2007.

\bibitem[Holliday et~al.(2007)Holliday, Chen, Tiampo, Rundle, Turcotte, and
  Donnellan]{holliday2007relm}
James~R Holliday, Chien-chih Chen, Kristy~F Tiampo, John~B Rundle, Donald~L
  Turcotte, and Andrea Donnellan.
\newblock A relm earthquake forecast based on pattern informatics.
\newblock \emph{Seismological Research Letters}, 78\penalty0 (1):\penalty0
  87--93, 2007.

\bibitem[Ward(2007)]{ward2007methods}
Steven~N Ward.
\newblock Methods for evaluating earthquake potential and likelihood in and
  around california.
\newblock \emph{Seismological Research Letters}, 78\penalty0 (1):\penalty0
  121--133, 2007.

\bibitem[Shen et~al.(2007)Shen, Jackson, and Kagan]{shen2007implications}
Zheng-Kang Shen, David~D Jackson, and Yan~Y Kagan.
\newblock Implications of geodetic strain rate for future earthquakes, with a
  five-year forecast of m5 earthquakes in southern california.
\newblock \emph{Seismological Research Letters}, 78\penalty0 (1):\penalty0
  116--120, 2007.

\bibitem[Kagan et~al.(2007)Kagan, Jackson, and Rong]{kagan2007testable}
Yan~Y Kagan, David~D Jackson, and Yufang Rong.
\newblock A testable five-year forecast of moderate and large earthquakes in
  southern california based on smoothed seismicity.
\newblock \emph{Seismological Research Letters}, 78\penalty0 (1):\penalty0
  94--98, 2007.

\bibitem[Ebel et~al.(2007)Ebel, Chambers, Kafka, and Baglivo]{ebel2007non}
John~E Ebel, Daniel~W Chambers, Alan~L Kafka, and Jenny~A Baglivo.
\newblock Non-poissonian earthquake clustering and the hidden markov model as
  bases for earthquake forecasting in california.
\newblock \emph{Seismological Research Letters}, 78\penalty0 (1):\penalty0
  57--65, 2007.

\bibitem[Diebold and Mariano(2002)]{diebold2002comparing}
Francis~X Diebold and Robert~S Mariano.
\newblock Comparing predictive accuracy.
\newblock \emph{Journal of Business \& economic statistics}, 20\penalty0
  (1):\penalty0 134--144, 2002.

\bibitem[Wickham(2016)]{ggplot}
Hadley Wickham.
\newblock \emph{ggplot2: Elegant Graphics for Data Analysis}.
\newblock Springer-Verlag New York, 2016.
\newblock ISBN 978-3-319-24277-4.
\newblock URL \url{https://ggplot2.tidyverse.org}.

\bibitem[South(2017)]{south2017rnaturalearth}
Andy South.
\newblock Rnaturalearth: world map data from natural earth.
\newblock \emph{R package version 0.1. 0}, 2017.

\bibitem[Ross(2016)]{ross2016bayesian}
GJ~Ross.
\newblock Bayesian estimation of the etas model for earthquake occurrences.
\newblock \emph{Preprint}, 2016.

\end{thebibliography}

\end{document}